\documentclass[aps,prb,reprint,showpacs,superscriptaddress,groupedaddress]{revtex4-1}
\usepackage{graphicx}
\usepackage{amsmath}
\usepackage{amssymb}
\usepackage{dcolumn}
\usepackage{latexsym}
\usepackage{rotating}
\usepackage[usenames,dvipsnames]{xcolor}
\usepackage{float}
\usepackage{epsfig}
\usepackage{psfrag}
\usepackage{natbib}
\usepackage{bm}
\usepackage{eucal}
\usepackage{mathrsfs}
\usepackage{braket}
\usepackage{enumerate}
\usepackage{longtable}
\usepackage{bm}
\usepackage{hyperref}
\usepackage{amsfonts}
\usepackage{dcolumn}
\usepackage{bm}
\usepackage{xspace}

\newcommand{\be}{\begin{equation}}
\newcommand{\ee}{\end{equation}}
\newcommand{\bn}{\begin{eqnarray}}
\newcommand{\en}{\end{eqnarray}}

\newcommand{\tise}{TiSe\textsubscript{2}\xspace}

\def\x2y2{{x^2-y^2}}

\usepackage{color} 

\usepackage{hyperref}
\hypersetup{
colorlinks=true,final=true,
        linkcolor=red,
        citecolor=blue,
        filecolor=blue,
        urlcolor=blue,
}
\begin{document}

\title{Lattice fluctuations, not excitonic correlations, mediated electronic localization in TiSe\textsubscript{2}}

\author{Ross E. Larsen}
\affiliation{Computational Science Center, National Renewable Energy Laboratory, 15013 Denver West Parkway, Golden CO 80401}
\affiliation{Renewable and Sustainable Energy Institute, University of Colorado Boulder, Boulder, Colorado 80309}
\author{Dimitar Pashov}
\affiliation{Theory and Simulation of Condensed Matter, King’s College London, The Strand, London WC2R2LS, UK}
\author{Matthew D. Watson}
\affiliation {Diamond Light Source Ltd, Harwell Science and Innovation Campus, Didcot, OX11 0DE, UK}
\author{Swagata Acharya}
\affiliation{Materials, Chemical and Computational Science Directorate, National Renewable Energy Laboratory, 15013 Denver West Parkway, Golden CO 80401}
\author{Mark van Schilfgaarde}
\affiliation{Materials, Chemical and Computational Science Directorate, National Renewable Energy Laboratory, 15013 Denver West Parkway, Golden CO 80401}
\email{Mark.vanSchilfgaarde@nrel.gov}

\begin{abstract}

TiSe\textsubscript{2} is thought to be an insulator with a bandgap of $\sim$0.1\,eV.  It has attracted a much
interest because, among of a rich array of unique properties, many have thought TiSe\textsubscript{2} is
a rare realisation of an excitonic insulator.  Below 200\,K, TiSe\textsubscript{2} undergoes a transition from a
high-symmetry ({$P\bar{3}m1$}) phase to a low-symmetry ({$P\bar{3}c1$}) phase. Here we establish that TiSe\textsubscript{2} is indeed an insulator in both {$P\bar{3}m1$} and {$P\bar{3}c1$} phases.
However, the insulating state is driven not by excitonic effects but by symmetry-breaking of the {$P\bar{3}m1$} phase.
In the CDW phase the symmetry breaking is static.  At high temperature, thermally driven
instantaneous deviations from {$P\bar{3}m1$} break the symmetry on the characteristic time scale of a phonon.  Even
while the time-averaged \emph{lattice} structure assumes {$P\bar{3}m1$} symmetry, the time-averaged \emph{energy band} structure is closer to the CDW phase -- a rare instance of a metal-insulator transition induced by dynamical symmetry breaking. We establish these conclusions from a high-fidelity, self-consistent form of many body perturbation theory, in
combination with molecular dynamics simulations to capture the effects of thermal disorder.  The many-body theory
includes explicitly ladder diagrams in the polarizability, which incorporates excitonic effects in an \emph{ab initio}
manner.  The excitonic modification to the potential is slight, ruling out the possibility that
TiSe\textsubscript{2} is an excitonic insulator.  Charge self-consistency is essential distinguish the
metallic from insulating state.

\end{abstract}
\maketitle

\section{Introduction}

TiSe$_{2}$ is a layered diselenide compound with space group $P\bar{3}m1$.  Below 200K, it undergoes a phase transition
to a charge density wave (CDW), forming a commensurate $2{\times}2{\times}2$ superlattice ($P\bar{3}c1$) of the original
structure.  At the transition there is a softening of the zone boundary phonon, and changes are seen in the transport
properties \cite{DiSalvo, Holt}.  It is also been observed that if the CDW is suppressed by
pressure~\cite{Kusmartseva09} or intercalation of Cu atoms~\cite{Morosan06}, (unconventional) superconductivity appears.
A quantum critical point has also been observed when pressure is applied~\cite{Abbamonte14}.  The interplay of the CDW
and superconductivity has been the subject of many studies, but even at the one-particle level whether the pristine
(undoped) system is a semimetal or insulator is not well understood.  A proper understanding of the origin of the
one-particle spectrum is a prerequisite for understanding the superconductivity, and forms the subject of this work.

Bianco et al. \cite{Mauri} discuss a controversy as to whether the instability originates either from the electrons or
by the lattice.  Such a distinction is not entirely well posed, because because nuclear fluctuations are correlative to
inverse of the static charge susceptibility \cite{Pick70}.  A less ambiguous way to frame the question is one of a
competition between a one-body effect, e.g. some analog of a Jahn-Teller mechanism
\cite{Motizuki80,Motizuki85,Motizuki87}, or a many-body effect, notably an excitonic insulator
\cite{Traum78,Wilson77,Cercellier}.  In their paper, Bianco et al. \cite{Mauri} present a detailed review of theoretical
and experimental understanding of the nature of the gap in the high temperature, $P\bar{3}m1$ phase.  It was established
early on~\cite{Zunger78} that the valence band maximum is of Se-4\emph{p} character and falls at the $\Gamma$ point,
while the conduction band minimum has Ti-3\emph{d} character and falls at L.  Experimental reports are voluminous and
somewhat contradictory, in part because the distinction between semimetal and a small-gap insulator is obscured by
doping.  A consensus has emerged, starting perhaps with the work of Ref.~\cite{Stoffel85} that the bandgap is small and
positive.  The gap is variously reported, in the range 0-0.2\,eV, and especially in earlier studies was reported as a
semimetallic phase with an inverted gap~\cite{Bachrach76,Zunger78,Chen80} in the high temperature phase.

The argument for TiSe\textsubscript{2} being an excitonic insulator largely stems from rather unusual shapes of the
spectral function observed by ARPES.  In the study of Ref.~\cite{Cercellier}, a large spectral weight from backfolded
bands was observed, which they interpreted as a signature of an excitonic insulator.  Ref.~\cite{Cazzaniga12} reported a
band of ``Mexican hat'' shape from a \emph{GW} calculation, also indicative of an excitonic insulator.  In this work we
indeed find backfolded bands, but they originate from Umklapp scattering processes, static in the CDW phase at 0K and
ephemeral at higher temperatures.  This will be discussed in more detail below.

Early calculations of the energy band structure found it to be a semimetal with the conduction band at L falling below
the valence band at $\Gamma$~\cite{Zunger78}, within the local-density approximation to density-functional theory (DFT).
It predicts a negative gap in both the {$P\bar{3}m1$} and CDW phases.  As the LDA systematically underestimates
bandgaps, it is not a reliable predictor of the gap.  Indeed, adding $U$ to the LDA, Bianco et al \cite{Mauri} found a
small opening of the gap in both {$P\bar{3}m1$} and CDW phases.  This is not determinative however, because LDA+\emph{U}
is ‘second-principles’ method whose answer depends \emph{U}, which is not known.  Prior to this work, Cazzaniga et
al. \cite{Cazzaniga12} considered a \emph{GW} calculation as a perturbation to the LDA, and found a gap of $\sim$200 meV
in the high-temperature {$P\bar{3}m1$} structure.  However, this result turns out to be an artefact of the
{$G^\mathrm{LDA}W^\mathrm{LDA}$} approximation, as we have reported previously~\cite{Acharya21a}.  Changes to the charge
density (which \emph{GW} as a first order perturbation to the LDA does not take into account) completely changes the
picture.  A Quasiparticle Self-Consistent form of \emph{GW} (QS\emph{GW})~\cite{mark06qsgw} predicts the {$P\bar{3}m1$}
structure to be semimetallic, with a gap approximately similar to the LDA though for different reasons.  In the LDA, not
only does the conduction band at L fall below the the VBM at $\Gamma$, but the Ti- and Se- derived levels at $\Gamma$
are also inverted, as can be seen from Fig.~2 of Ref.~\cite{Acharya21a}.

\section{Results and Discussion}

Since QS\textit{GW} is a much higher fidelity theory~\cite{mark06adeq} than $G^\mathrm{LDA}W^\mathrm{LDA}$, and moreover
it usually tends to slightly overestimate insulating bandgaps, it is unlikely that ideal {$P\bar{3}m1$}
TiSe\textsubscript{2} is an insulator.  It is highly unusual that {$G^\mathrm{LDA}W^\mathrm{LDA}$} predicts a gap while
QS\textit{GW} does not, but it can be explained in terms of renormalisation of the quasiparticle band structure owing to
corrections to the LDA density~\cite{Acharya21a}.  Moreover it offers a hint that excitons are indeed responsible for
forming a gap, since excitonic effects are absent from \textit{GW}.  However, as we show here, this is not the case.
The role of excitonic effects can be assessed in an \textit{ab initio} framework by extending \textit{GW} to include
ladder diagrams in \textit{W} ($W{\rightarrow}\widehat{W}$)~\cite{Cunningham2023}.  We find that electron-hole contributions
negligibly affect the self-energy.  Nevertheless TiSe\textsubscript{2} is a band insulator in both {$P\bar{3}m1$} and
CDW phases, but the gap forms only when nuclear fluctuations are properly taken into account.  We establish it in the
next section by combining {\it ab initio} molecular dynamics (AIMD) simulations with QS\textit{GW} calculations of the
band structure on configurations sampled from the AIMD trajectories.

\subsection{The Charge-Density Wave}

Below 200\,K, Di Salvo et al\cite{DiSalvo} found from neutron diffraction experiments that TiSe\textsubscript{2}
reconstructs into a ``$3q_\mathrm{L}$'' charge-density wave (CDW).  In the CDW phase ({$P\bar{3}c1$} symmetry), the
three-atom {$P\bar{3}m1$} unit cell undergoes a $2{\times}2{\times}2$ reconstruction into a 24-atom superlattice.  The
$3q_\mathrm{L}$ superstructure is realised by superimposing on $P\bar{3}m1$ displacements from a symmetric linear
combination of three L point phonon modes.  L points fall at a zone boundary: they are projected onto the $\Gamma$ point
under a $2{\times}2{\times}2$ reconstruction.  Thus the CDW is commensurate with a $2{\times}2{\times}2$ superlattice of
the $P\bar{3}m1$ structure, with $P\bar{3}c1$ symmetry.

Bianco et al~\cite{Mauri}, used the GGA to estimate in detail the nuclear positions of this 24-atom structure.  They
tried several kinds of constraints (van der Waals interactions play an important role, which the GGA is known not to
treat well), in particular consider a scenario in which the lattice vectors are fixed at experimental values and nuclear
positions relaxed under that constraint.  The resulting equilibrium geometry is very near the experimentally observed
$3q_\mathrm{L}$ distortion of the $P\bar{3}m1$ phase, in line with Di Salvo's experiments.  Initially we will adopt the
experimental structure, as it is the most reliable.

\begin{table}[h]
\caption{\label{tab:superlattice} Linear combinations of primitive lattice vectors \textbf{T} in the $P\bar{3}m1$ cell that
  comprise the $P\bar{3}c1$ supercell, and combinations of the corresponding primitive reciprocal lattice
  vectors \textbf{G} which make up the \textbf{k} points that fold into the $\Gamma$ point.
}
\begin{ruledtabular}
\begin{tabular}{l|lll|cccl}
         & $\mathbf{T}_{1}$ & $\mathbf{T}_{2}$ & $\mathbf{T}_{3}$ & $\mathbf{G}_{1}$ & $\mathbf{G}_{2}$ & $\mathbf{G}_{3}$ \\
\hline
$\Gamma$ & 0 & 0 & 0 \quad  & 0   & 0   & 0   & \quad\\
A     	 & 0 & 0 & 1 \quad  & 0   & 0   & 1/2 & \quad\\
M$_{1}$  & 1 & 0 & 0 \quad  & 0   & 1/2 & 0   & \quad\\
M$_{2}$  & 0 & 1 & 0 \quad  & 1/2 & 0   & 0   & \quad\\
M$_{3}$  & 1 & 1 & 0 \quad  & 1/2 & 1/2 & 0   & \quad\\
L$_{1}$  & 1 & 0 & 1 \quad  & 0   & 1/2 & 1/2 & \quad\\
L$_{2}$  & 0 & 1 & 1 \quad  & 1/2 & 0   & 1/2 & \quad\\
L$_{3}$  & 1 & 1 & 1 \quad  & 1/2 & 1/2 & 1/2 & \quad\\
\end{tabular}
\end{ruledtabular}
\end{table}

Ref.~\cite{Mauri} explained in some detail how to construct displacements of individual normal modes that form a $3q$
CDWs and we recapitulate their description here.  TiSe$_{2}$ is a layered compound with Ti lying in one layer in the
basal plane, sandwiched between a pair of Se layers above and below.  The $P\bar{3}m1$ structure has three independent
parameters: lattice parameters $a$ and $c$, and the height $h$ of the Se plane relative to Ti.  The triplet of (Se$_1$,
Ti, Se$_2$) layers repeats periodically along the $c$ axis.  $P\bar{3}c1$ is comprised of 8 unit cells of $P\bar{3}m1$
(doubling the lattice along each axis); thus 8 \textbf{k}-points fold into the $\Gamma$ point of the larger cell.  These
points are depicted in Table~\ref{tab:superlattice}, expressed as multiples of primitive lattice vectors of
$P\bar{3}m1$.  A phonon mode of L$_{1}$ symmetry will have equivalent modes at L$_{2}$ and L$_{3}$, related by a
three-fold rotation about the $c$ axis.  More details concerning $3q_{L}$ structure can be found in the Appendix.

A great deal can be learned by considering the evolution of the CDW along a generalised coordinate $x$ where

\begin{eqnarray}
\mathbf{u}_i = \mathbf{u}_i^0 + x (\mathbf{u}_i^{3q_L} - \mathbf{u}_i^0)
\end{eqnarray}

$\{\mathbf{u}_i^0\}$ and $\{\mathbf{u}_i^{3q_L}\}$ are coordinates respectively of the $2{\times}2{\times}2$
superlattice of the $P\bar{3}m1$ and the $3q_L$ CDW.  Coordinates are taken from the supplemental material of
Ref.~\onlinecite{Mauri}; we use the experimental lattice parameters and displacements ($\nu$=3), and choose the overall
amplitude so that the lattice reaches experimental equilibrium geometry at $x=x^*=1$, i.e. when the L$_{1}$ component of
$|\mathbf{u}|_\mathrm{Se}$=0.014\,\AA\ (see Eq. 10 of Ref.~\onlinecite{Mauri}).  The displacements are quite small:
at $x{=}x^*$) a typical change in bond length is only $\delta d/d \approx 0.02$.

\begin{figure}[h!]
\centering
\includegraphics[height=2.9cm]{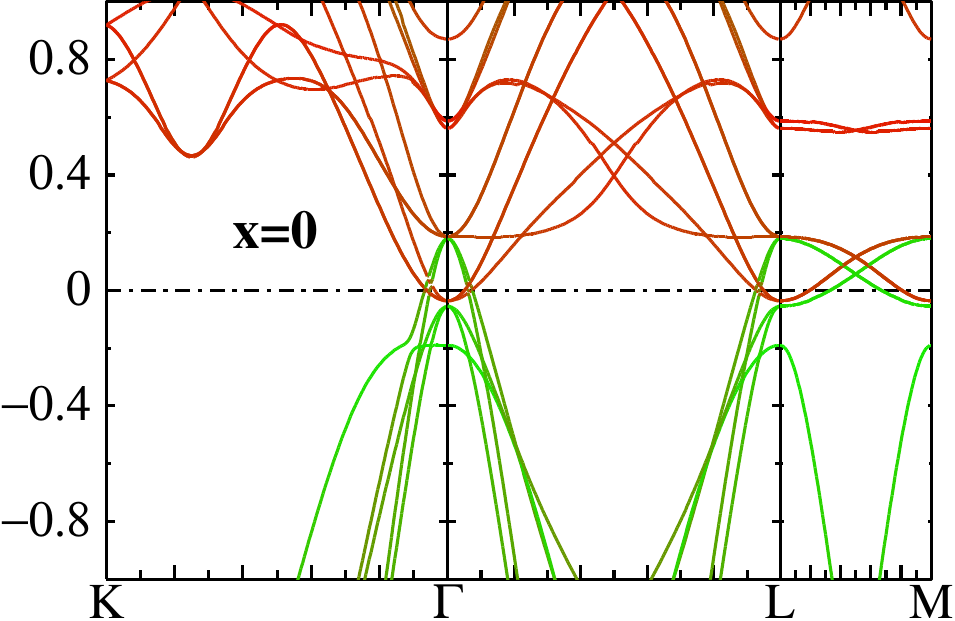}\
\includegraphics[height=2.9cm]{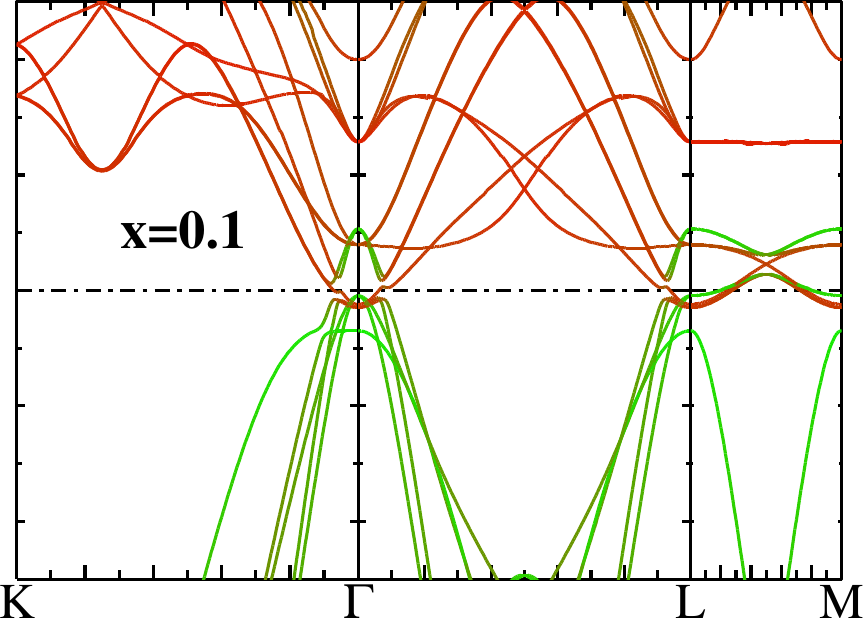}
\includegraphics[height=2.9cm]{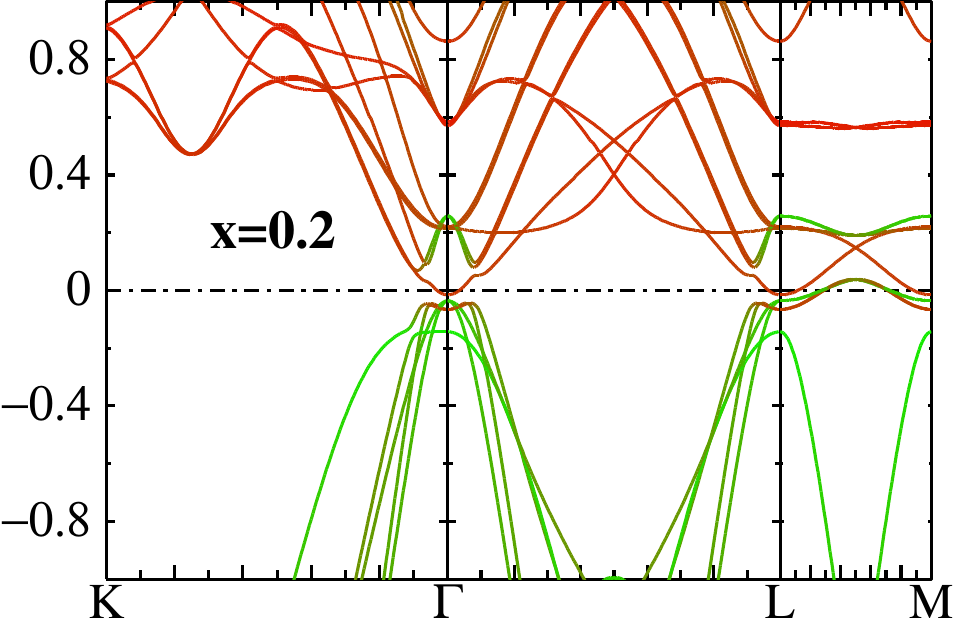}\
\includegraphics[height=2.9cm]{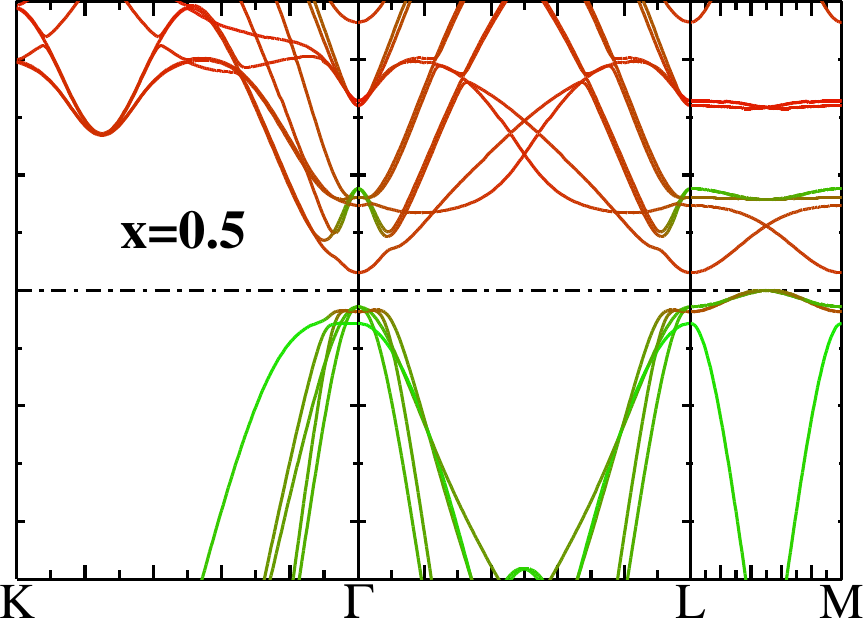}
\includegraphics[height=2.9cm]{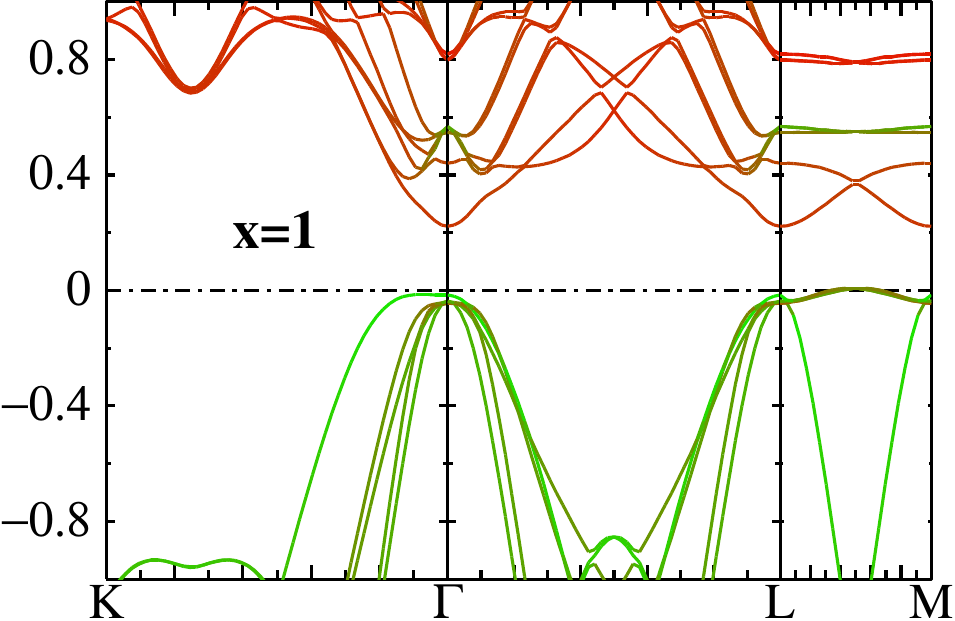}\
\includegraphics[height=2.9cm]{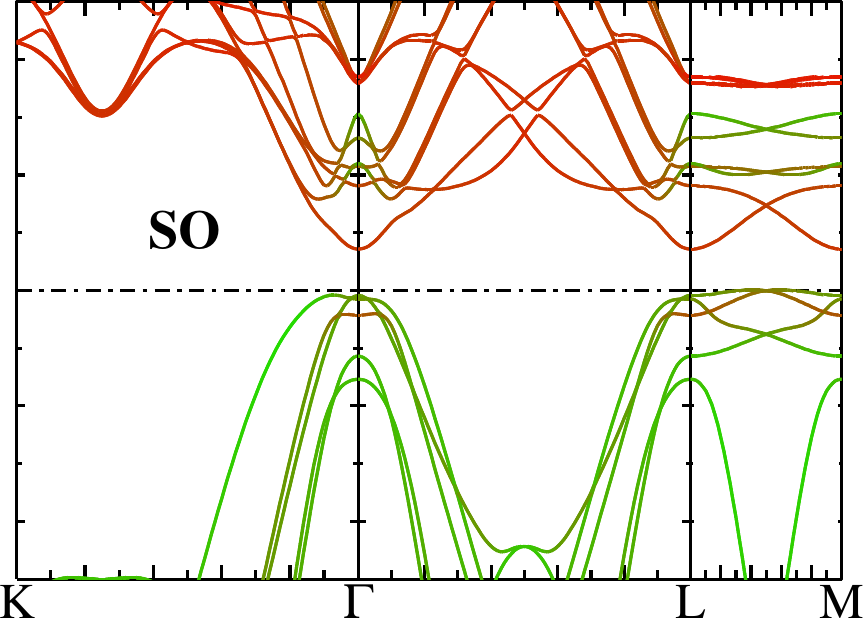}
\caption{QSGW Energy bands of the $P\bar{3}c1$ superlattice for:
  $(a)$ the undistorted structure, $x$=0;
  $(b)$ the initial stages of the CDW ($x$=0.1);
  $(c)$ formation of gap on $\Gamma$-K and $\Gamma$-L lines ($x$=0.2);
  $(d)$ fully formed metal-insulator transition ($x$=0.5);
  $(e)$ band structure at theoretical equilibrium point ($x$=1);
  $(f)$ band structure at $x$=1 with spin-orbit coupling included.
  Red and green indicate a projections onto Ti-3\textit{d} and Se characters.
  Note that L, M, and K are half the size of L, M, and K in the {$P\bar{3}m1$} phase.  The latter points fold into
  $\Gamma$ in the CDW.
}
\label{fig:fig2}
\end{figure}

Fig.~\ref{fig:fig2} shows how the band structure evolves with the initiation of the CDW.  Panel \ref{fig:fig2}$(a)$
merely reproduces the band structure of the high-symmetry {$P\bar{3}m1$} structure, in the $2{\times}2{\times}2$
superstructure which folds L and M into $\Gamma$.  Thus the lines $\Gamma{\rightarrow}\mathrm{L}$ and 
$\mathrm{L}{\rightarrow}\mathrm{M}$ connect equivalent points along different directions.
There is no coupling between L-derived states near $E_{F}$ (mostly Ti character, red) and $\Gamma$-derived states
(mostly Se character, green) owing to translational symmetry of the smaller $P\bar{3}m1$ structure.  The Ti-like $d$
states at $\Gamma$ ($E_{F}-0.04$\,eV) are triply degenerate; the Se-like $p$ states at +0.2\,eV are doubly degenerate.
Note in particular a pair of doubly degenerate Se bands at $E_{F}{-}0.05$\,eV and a corresponding pair at
$E_{F}{+}0.18$\,eV.  Midway along the $\Gamma$-M line (which we denote Q for convenience) they cross and form a fourfold
degeneracy at the midpoint.  Something similar occurs with pairs of threefold degenerate Ti-derived bands at
$E_{F}{-}0.04$\,eV and $E_{F}{+}0.18$\,eV.  For a gap to appear these 10 states must split into a group of four occupied
and six unoccupied states.

With the initiation of the CDW (see Fig.~\ref{fig:fig2}, $x$=0.2), two pairs of L-derived (Ti-like) states near $\Gamma$
begin to admix with $\Gamma$-derived (Se-like) states.  They repel and form a gap near $\Gamma$, with the band maximum
offset from $\Gamma$.  The third Ti-like state at $E_{F}-0.04$\,eV begins to separate from other two.  Eventually it
rises enough to completely separate from the valence bands and form the conduction band minimum.  However the splitting
increases quadratically with $x$, so for $x{<}0.15$ it overlaps with the two aforementioned bands and no gap is formed.
Around $x{=}0.2$ there is a pseudogap: a gap on the $\Gamma$-K and $\Gamma$-L lines is fully formed, but the system
remains metallic on the L-M line.  For a gap to form around Q, the aforementioned pair of doubly degenerate Se-like
states ($-$0.05 and +0.2\,eV at the endpoints) that cross midway between L and M also couple to form bond-antibond
pairs.  Each of the three kinds of splittings (Se-Se splitting at Q, Ti-Se coupling near $\Gamma$, Ti-Ti splitting at
$\Gamma$) conspire together to eventually form a gap for sufficiently large $x$.  The gap is fully formed when $x$=0.5; however,
it is indirect with the valence band maximum at Q and conduction band minimum at $\Gamma$.

As $x$ increases further to the equilibrium point at $x{=}1$ the topology remains similar, and the gap continues to
widen.  Spin-orbit coupling changes the topology somewhat and reduces the gap to about 0.14\,eV.  The topology of the
valence bands is remarkable with several distinct points all within $kT$ of each other.  The highest band along L-M
(nominally the valence band maximum) almost completely dehybridizes and appears like a flat, atomic-like state.  The
colours also show that a state at $\Gamma$ below the VBM has Ti-3\textit{d} character.  This is an echo of the L-derived
conduction band minimum in the high-symmetry {$P\bar{3}m1$} phase.  The evolution of the gap with respect to $x$ makes
it clear that the gap opens as a consequence of level repulsions in the CDW superlattice that appear because states of
unique wave number in the erstwhile {$P\bar{3}m1$} Brillouin zone become coupled.  This is in keeping with the
observation of Stoffel et al.~\cite{Stoffel85}, and of Cercellier et al.~\cite{Cercellier} who directly observed
zone-folding effects in ARPES measurements.  

\noindent\textit{Excitonic effects}

\begin{figure}[h!]
\centering
\includegraphics[height=3.5cm]{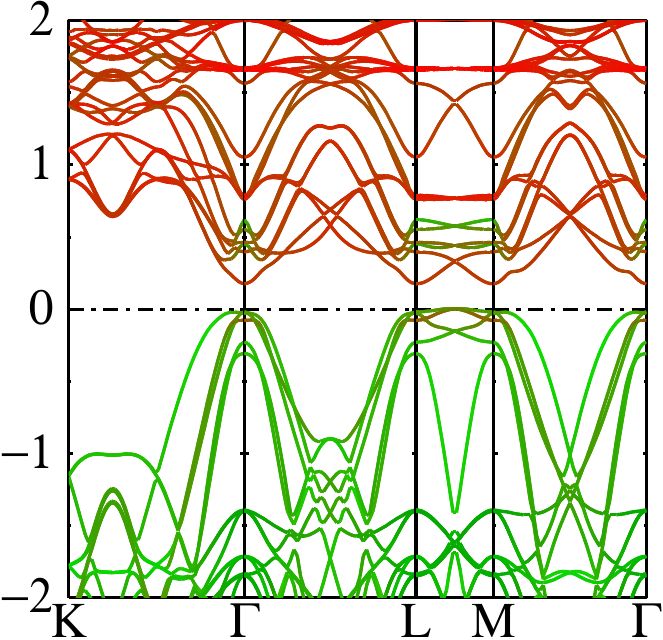}
\includegraphics[height=3.5cm]{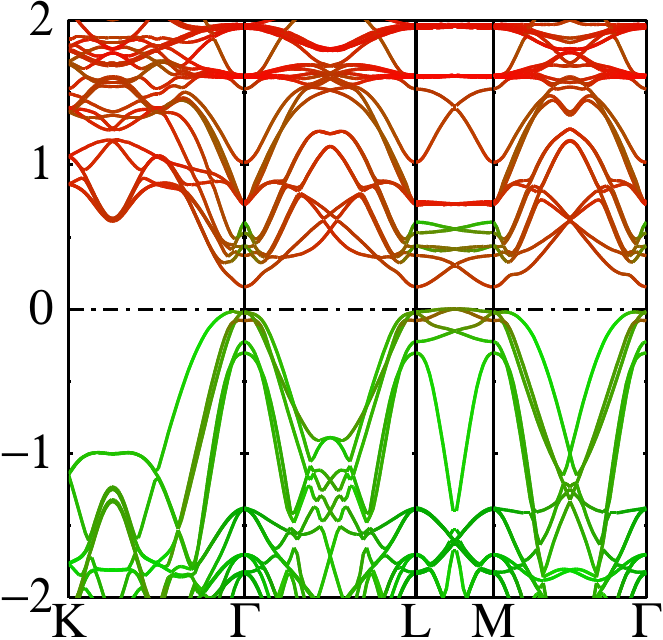}
\caption{Energy bands of the $P\bar{3}c1$ superlattice in the QS\textit{GW} approximation (left) and QS$G\widehat{W}$
  approximation (right), respectively, in the Brillouin zone corresponding to the {$P\bar{3}m1$} phase.  Red indicates a
  projection onto Ti-3\textit{d} character, green onto Se character.
}
\label{fig:bse}
\end{figure}

We can incorporate electron-hole effects in the \textit{GW} approximation by adding ladder diagrams to the polarizability
via a Bethe-Salpeter equation.  We use QS$G\widehat{W}$ to indicate an enhancement of QS$GW$ when ladder diagrams are
added to \textit{W}.  Such a capability was recently developed within Questaal, and Ref.~\cite{Cunningham2023} examines in
detail the change in spectral properties when $W$ is augmented to $\widehat{W}$ for a wide range of materials systems.
It was shown to significantly improve the fidelity of the QS$GW$ theory, particularly when correlations are weak or
moderate, as is the case with TiSe\textsubscript{2}.  Almost universally, both one-particle spectral functions and
dielectric functions are described with remarkably high fidelity.

Thus QS$G\widehat{W}$ provides a rigorous, \textit{ab initio} framework to establish the role excitons play in
TiSe\textsubscript{2}, and whether it might be an excitonic insulator.  Fig~\ref{fig:bse} compares the QS$GW$ and
QS$G\widehat{W}$ energy band structures in the CDW phase.  Differences are very slight: the ladder diagrams reduce the
fundamental gap by 0.02\,eV.  The weak influence of excitons on spectral functions is typical of narrow-gap systems: the
nearly metallic electronic structure causes the RPA part of the polarizability to be large, and ladders act to make a
small correction to it.

The twin observations that QS$GW$ without excitonic effects yields a small gap (which agrees with experiments as well as
they can be measured), and that excitonic additions have almost negligible effect, lead us to conclude with some
confidence that TiSe\textsubscript{2} is not an excitonic insulator in the CDW phase.

\subsection{Structural fluctuations at finite temperature}

Above the ordering temperature $T_{\mathrm{SDW}}{\sim}200$\,K, the CDW is lost and the system changes to the
high-symmetry $P\bar{3}m1$ phase.  However, it assumes $P\bar{3}m1$ only when averaged over time: nuclei are not fixed
but fluctuate stochastically about the ideal lattice positions.  At very low temperature excursions will be distributed
as harmonic excursions around one of the several possible $3q_L$ deformations to the CDW.  Above the transition
temperature the lattice assumes the $P\bar{3}m1$ structure on average but will undergo fluctuations which will be
weighted to a locale resembling one of the $3q_L$ CDWs.  What effect this has on the energy-band structure can only be
determined from knowledge of the stochastic motion of the nuclei.

To this end we performed classical \textit{ab initio} molecular dynamics AIMD simulations of a 96-atom simulation cell at the
density-functional (DFT) level, and took selected snapshots separated far enough in time so as to be uncorrelated.  This
amounts to making an adiabatic approximation for the electronic structure, and follows a strategy similar to that used successfully to determine the
dielectric response and temperature dependence of the bandgap renormalisation in Si~\cite{Zacharias16}.  For context,
{$P\bar{3}c1$} has 3 atoms in the unit cell: the 96-atom cell corresponds to a $4{\times}4{\times}2$ replication of it,
and a $2{\times}2{\times}1$ superlattice of the CDW.  The 24-atom CDW is the smallest system capable of revealing
differences above and below the CDW phase transition; however, to allow for additional sampling of thermal fluctuations
on length scales larger than supported with 24 atoms, we focused on the 96 atom cell. More complete details of how we performed the MD simulations are
given Section \ref{aimd_method}.

As we show below, the gap is indeed preserved at finite temperature.  This is a rare instance when
\textit{thermal nuclear fluctuations qualitatively change} the topology of the Fermi surface, from metal to insulator.  These
are largely a result of Umklapp processes that couple $k$ points near $\Gamma$ to those near L.  While the effect of
nuclear fluctuations on the bandgap has been noted in other systems~\cite{Etienne16,McKechnie18}, TiSe\textsubscript{2}
is unique in that the Umklapp processes give rise to a metal-insulator transition.  These Umklapp processes arise
because of dynamical Peierls-like displacements that result from the ``softness'' of the system, which follow as a
consequence of the internal energy being more favourable for lattices deformed relative to the $P\bar{3}m1$ structure. 

\subsection{Electronic structure from molecular dynamics trajectories}
\label{sec:bands96}

\noindent\textit{Molecular dynamics simulations}

\begin{table}[h]
\begin{tabular}{c|c c|c c|c c}\hline\hline
\vbox{\vskip 12pt}
          & $\overline{d}$ & $\Delta{d}$ & $\overline{\theta}$ & $\Delta{\theta}$ & $\overline{E}_G$ & $\Delta{E}_G$  \cr
{$P\bar{3}m1$} (Expt)
          & 2.554          & -           & 90.0                &  2.3             & $\sim$0.1        &   -   \cr
CDW(Expt) & 2.556          & 0.038       & 90.0                &  2.5             & 0.18             &   -   \cr
120\,K    & 2.564          & 0.067       & 90.0                &  3.7             & 0.09             &  0.05 \cr
300\,K    & 2.568          & 0.089       & 89.9                &  4.4             & 0.14             &  0.06 \cr
\hline
\end{tabular}
\caption{Data extracted for the {$P\bar{3}c1$} in the experimental geometry~\cite{DiSalvo}, and from molecular dynamics simulations in the the 96 atom cell at 120\,K and 300\,K.  
  Columns correspond to mean nearest-neighbour Ti-Se bond length $\overline{d}$ and fluctuations about the mean (\AA);
  mean Se-Ti-Se bond angle $\overline{\theta}$ (degrees) and fluctuations about the mean, and the mean QS$GW$ bandgap $\overline{E}_G$
 (eV) and root-mean-squared fluctuations about the mean.}
\label{tab:mdstats}
\end{table}

Details on how the AIMD were carried out are given in a later section~\ref{aimd_method}.  Table~\ref{tab:mdstats} shows
that average bond lengths and angles are similar in the {$P\bar{3}c1$} and {$P\bar{3}m1$} phases, so it would be
difficult to distinguish them based on pair distribution functions. Indeed, the thermally broadened
radial distribution functions look similar at both 120 K and 300 K, albeit with slightly greater broadening of peaks at 300 K.

We computed the QS$GW$ band structure from twelve snapshots calculated at nuclear configurations generated from the
AIMD, at 120\,K and 300\,K.  Band edge states are quite fluid in both conduction and valence bands: the valence band
maximum fluctuates between the $\Gamma$ and A points.  At either temperature the gap varies between 0 and 0.25\,eV, with RMS
fluctuation of $\sim$0.06\,eV (Table~\ref{tab:mdstats}). As for the disagreement between the absolute values of the gap between theory and experiments ($\sim$60 meV), it is worth noting that realistically, experiments are performed with samples that are degenerately n-doped with $\sim$1$\%$ excess electrons, and are not stoichiometric. In such situations the change in the gap values between the two phases is a more physical quantity to consider.  An  intriguing observation is the reduction of the gap at lower temperatures. We note that the lower bound of our estimated gap values at 300 K is $\sim$80 meV which is close to the experimental observation of $\sim$74 + $\pm$15 mev gap at the same temperature~\cite{PhysRevLett.122.076404}. At 120 K, we estimate a gap of 90 meV $\pm$ 60 meV of RMS fluctuations when the QS$GW$ gap values are sampled over all MD snapshots, i.e. is a reduction of about 50 meV in the mean gap value from 300 K. This is in remarkable agreement with the estimated experimental reduction in gap of 60 meV inside the CDW phase (experimentally estimated gap at 10 K is 15 meV)~\cite{PhysRevLett.122.076404}. Note that our calculations are performed at 120 K while the experiments are performed at 10 K, and the difference in temperature could account for $\sim$10 meV in thermal broadening.  

That increasing thermal lattice fluctuations cause the electronic gap to widen is intriguing. How itinerant electrons localize is at the heart of microscopic principles of electronic correlations, as well as devices based on localisation. For a large class of semiconductors, electrons are Bloch states that open up a gap at the Fermi energy due to reflections at the Brillioun zone boundary, a phenomenon that is known as metal to band insulator transition. Most non magnetic semiconductors are trivial band insulators where the emergence of the insulating phase has a one-to-one correspondence with Landau Fermi liquid theory and for all such cases the electronic properties can be described by single Slater determinant. However, for Mott insulators, magnetic fluctuations, in particular, a multi-reference scenario (multiple Slater determinants) are ubiquitous in describing the localization of electrons starting from the metallic phase and formation of a gap, a phenomenon that has no analogue in Landau Fermi liquid theory for electronic excitation. However, once the electrons get localized and the gap at the Fermi energy is formed, phonons, often than not, play a secondary role of reducing the band gaps~\cite{Etienne16,McKechnie18} by enhancing the electronic screening. In solid state systems, it is  rare to find an occasion where electronic localization in itself is mediated by phonons, making TiSe$_{2}$ a special case. It is in that sense that the electronic localization is TiSe$_{2}$ is more akin to how localization happens in liquids. In TiSe$_{2}$ the lattice remains significantly soft and squeezy over a large range of temperatures, so much so, that it can be a source of disorder in itself that is sufficient to localize electrons.  

\begin{figure}[h!]
\centering
\includegraphics[height=3.5cm]{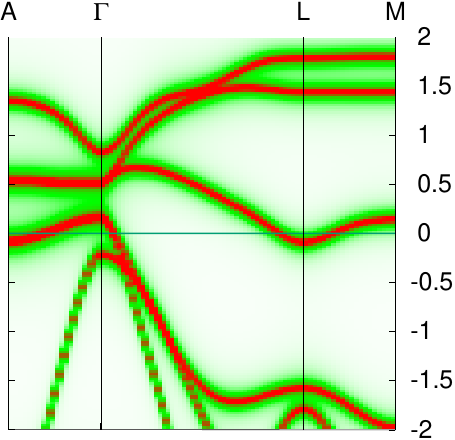}\
\includegraphics[height=3.5cm]{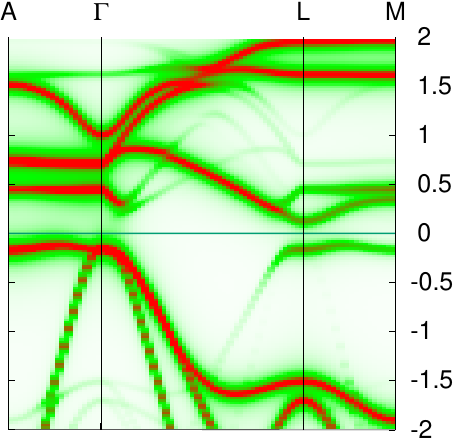}
\includegraphics[height=3.5cm]{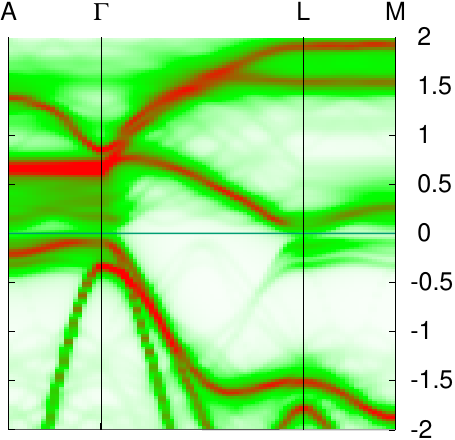}\
\includegraphics[height=3.5cm]{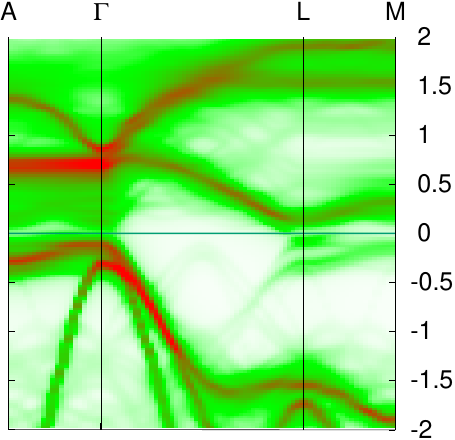}
\caption{QS$GW$ spectral functions folded to the Brillouin zone of the {$P\bar{3}m1$} structure for $(a)$ the ideal
  {$P\bar{3}m1$} structure, $(b)$ the CDW phase, $(c)$ the statistically averaged snapshots from the 120\,K AIMD
  simulations and $(d)$ the same for the 300\,K simulations.  The color scale ranges from no intensity (white) to full intensity (red) with green midway between.}
\label{fig:zonefoldedbands}
\end{figure}

To recover an effective band structure in the {$P\bar{3}m1$} phase, we use a band unfolding technique along the lines of
work presented in Refs.~\cite{Dargam97,Medeiros14}.  Zone unfolding allows us to represent $k$-resolved spectral
information of the $4{\times}4{\times}2$ supercell as statistically averaged quantities of the {$P\bar{3}m1$} Brillouin
zone.  The Brillouin zone of the primitive cell is folded into the supercell so that $N$ \textbf{k}-points in it
$\mathbf{k}_1,\dots,\mathbf{k}_N$, become reciprocal lattice vectors $\mathbf{G}_1,\dots,\mathbf{G}_N$ of the supercell
($N{=}24$ in this case), with one of the $\mathbf{G}$ vectors being $\mathbf{G}{=}0$.  While details of our
implementation will be presented elsewhere, in brief it is a method to resolve an eigenstate ${\psi}^{n\mathbf{k}}$ of
the supercell into linear combinations of eigenstates
$\widetilde{\psi}^{n'\mathbf{k}_1}\dots\widetilde{\psi}^{n'\mathbf{k}_N}$ of the primitive cell.  Since ${\psi}^{n\mathbf{k}}$
is normalised, it will be distributed among partial contributions from the
$\widetilde{\psi}^{n'\mathbf{k}_1}\dots\widetilde{\psi}^{n'\mathbf{k}_N}$.  The
partial weights are the scattering amplitudes into these states from Umklapp processes.
The $\mathbf{G}{=}0$ weight for eigenstate $w_{\mathbf{k}}^n$ is less than unity, approaching it when $\mathbf{k}$ is a good
quantum number, i.e. if the supercell is merely a periodic replica of the primitive cell.  Thus we can express the
spectral function as
\begin{align}
A(\mathbf{k},\omega) = \sum_{i} w_{\mathbf{k}}^i\, \delta[\omega-(\varepsilon_i-E_F)]
\label{eq:sfun}
\end{align}
where $i$ runs over the eigenstates of the supercell at \textbf{k}.  Eqn.~\ref{eq:sfun} provides a reasonable
description for $A(\mathbf{k},\omega)$ in the Brillouin zone of the primitive cell, for any snapshot, but averaging over
snapshots gives a better representation.  The result is shown in Fig.~\ref{fig:zonefoldedbands}.  The poles in
Eq.~\ref{eq:sfun} were broadened out by a Lorenzian broadening of 20\,meV.

Fig.~\ref{fig:zonefoldedbands}(a) shows the energy bands of the ideal {$P\bar{3}m1$} structure.
\textbf{k{}} is a good
quantum number and bands are well defined bands without Umklapp processes.  Fig.~\ref{fig:zonefoldedbands}(b) shows how
Umklapp processes open a gap in the CDW.  A replica of the highest valence band on the A--$\Gamma$ line appears on the
L--M line.  This couples to the Ti-$3d$ state on the L--M line, forming one bonding state and two antibonding states,
thus opening a gap.  Similarly, the Ti-$d$ state at L picks up a replica at $\Gamma$; indeed the majority of the QP
weight at L is transferred to $\Gamma$. Turning to the 120\,K case, $A(\mathbf{k},\omega)$ falls between the
{$P\bar{3}m1$} and CDW limiting cases.  The Umklapp processes are weaker but nevertheless present, causing a gap to open
very similar to the CDW case.  The highest valence band on the A-$\Gamma$ line is almost dispersionless, with
significant incoherence, and very likely these factors are the driving force for many-body effects such as
superconductivity in doped TiSe\textsubscript{2}.  At 300\,K, Umklapp processes are weaker but still present and the gap
is preserved.  This is a remarkable result, since only some kinds of deformations are sufficient to open a gap (see
Fig.~\ref{fig:fig2}).  Intriguingly enough, the blob of spectral weight at L (which is missing in the ideal {$P\bar{3}m1$} structure but is only there when the lattice fluctuations are taken into account), can be observed in experiments too~\cite{PhysRevB.81.155104}.  At low temperatures (deep inside the CDW phase) one obviously sees the strongly backfolded hole bands. However at temperatures immediately above 200 K (outside the CDW phase), one can still clearly observe this blob of spectral weight with similar intensity as that of the CDW phase that gradually disappears with increasing temperatures and by 300 K it is mostly gone. The fact that this blob of spectral weight at and above 200 K persists and is temperature dependent are strong signatures that these are related to the fluctuations of the $2{\times}2{\times}2$ CDW order. However, it is only natural that the exact nature of the coherence of these blobs are sensitive to the de-phasing that comes from the finite size effects in theory. The larger the supercell is higher is the de-phasing and more quickly the blobs will be washed out as we go further away and above the CDW temperature. 

An intriguing aspect of systems with an ordered phase is the fluctuation of the order parameter that persists at much higher temperatures compared to the critical temperature for the ordered phase. Experimental methods can barely measure the fluctuation itself above the ordered phase and often the information of the fluctuations are inferred from indirect microscopic signatures that reflect in various measurements. A study of these indirect microscopic signatures in various materials is of extreme interest. Knowledge of the parameter space (temperatures and pressure) where these fluctuations survive is key to finding experimental parameters that can help in condensing the fluctuations into ordered phases at different temperatures and pressures. For example, the experimental signatures of preformed Cooper pairs in cuprates~\cite{PhysRevX.11.031068,seo2019evidence,bovzovic2020pre,maki1968critical}, iron based superconductors~\cite{PhysRevLett.125.097003}, organic superconductors,~\cite{PhysRevResearch.5.023165} CDW fluctuations in various chacogenides~\cite{feng2012order} and cuprates~\cite{arpaia2019dynamical} and beginning of formation of Kondo cloud~\cite{v2020observation} have been matters of intensive studies in the recent years. Often, multiple orders and their associated order-parameter fluctuations are present in these materials at a certain part of the phase diagram and these fluctuations can compete, co-operate or coexist. In such situations it becomes difficult to attribute certain microscopic signatures to the fluctuations associated with a particular order parameter. A similar situation occurs in TiSe$_{2}$ where (since the spin fluctuations are absent) the charge fluctuations (particularly the excitonic correlations) and lattice fluctuations can leave their imprint on various measurements. Our work is a step change in this regard, where we unambiguously determine the signature of the CDW order parameter fluctuations that persist upto 300 K and show that the spectral blobs observed in the ARPES can not be due to the excitonic correlations but fully determined by the lattice fluctuations. The additional presence of the magnetic fluctuations in various cuprates, iron base supercondcutors make the analysis more involved, but a systematic diagrammatic improvement of our many-body perturbative approach is probably the desired way forward in addressing the outstanding problem in those class of magnetic materials too.

\section{Computational Methods}

\subsection{Band structure and QSGW}\label{qsgw}

The Quasiparticle Self-Consistent \textit{GW} form of \textit{GW}~\cite{mark06qsgw,Kotani07} dramatically improves its
fidelity compared to the usual (DFT-based) \textit{GW}.  Details of the theory and its implementation in
Questaal~\cite{questaal_web} can be found in Refs~\cite{Kotani07,questaal_paper}.  QS\textit{GW} largely eliminates the
starting-point dependence (a problem for \textit{GW} implementations) and importantly, errors are highly systematic,
which clarifies which diagrams are missing.  In the spectrum between Hartree Fock (which overestimates gaps) and DFT
(which understimates them) QS\textit{GW} has a tendency to err slightly on the HF side.  As we have shown
recently~\cite{Cunningham2023}, adding ladder diagrams to the polarizability eliminates the tendency to underscreen
greatly reduces this error.  As a consequence excited states and optical properties are shown to be predicted with high
fidelity, particularly if correlations are modest as they are here.  As we have shown, the high fidelity of
QS\textit{GW} and QS$G\widehat{W}$ is essential to distinguish artefacts of approximations in the theory from physical
behaviour of TiSe\textsubscript{2}.

\subsection{Ab initio molecular dynamics}\label{aimd_method}

All AIMD simulations were performed at the density functional theory (DFT) level using the Vienna Ab-Initio Simulation
Package (VASP 5.4)\cite{kresse1993ab,kresse1994ab,kresse1996efficiency,kresse1996efficient} with the
Perdew-Burke-Ernzerhof (PBE) exchange correlations functional\cite{perdew1996generalized} and the associated projector
augmented wave (PAW) pseudopotentials\cite{blochl1994projector,kresse1999ultrasoft} with a plane-wave energy cutoff of
400 eV and periodic boundary conditions in all three directions. A $3{\times}3{\times}3$ $\Gamma$-centered k-point grid
was used, and Grimme's D3 van der Waals dispersion-energy correction (DFT-D3)\cite{grimme2010consistent} was applied
with a timestep of 2 fs. Fermi smearing was applied with a smearing energy of 0.026 eV (consistent with a temperature of
300 K; for consistency this was used at 120 K as well).

We determined the minimum size of k-point mesh and the choice of van der Waals correction for the AIMD by taking the 24 atom CDW cell and allowing both the atomic positions and the lattice dimensions to relax to minimize the energy, for several different combinations of k-space mesh and van der Waals dispersion correction. We chose a combination of mesh and dispersion correction that closely matches the experimentally determined cell volume of 521.47 \AA. The results of these relaxations are  shown in Table \ref{kmesh_and_vdw}. Relaxation of the lattice with the Grimme DFT-D2 and DFT-D3 corrections and $2\times 2\times2$ and $3\times3\times3$ meshes gives geometries close to the experimental value. The DFT-D3 van der Waals dispersion correction was selected because it gives volumes slightly smaller than the experimental volume, which is consistent with there being some contraction from relaxation (essentially a zero temperature result). The $3\times3\times3$ k-space mesh (bolded) was selected because it is the smallest tested mesh that contains the $\Gamma$ point and gives a volume within 0.5\% of the experimental volume.

\begin{table}
\begin{tabular}[b]{|c|c|c|c|c|c|} 
\hline
VDW & $\Gamma$ & $2\times2\times2$ & $3\times3\times3$ & $4\times4\times4$ & $5\times5\times5$ \cr\hline
None & 683.72 & 560.87 & 562.65 & 556.04 & 572.91 \cr
TS$^a$ & 619.11 & 502.25 & 500.13 & 499.04 & 500.49 \cr
TS-H$^b$ & 629.93 & 507.10 & 507.20 & 503.93 & 505.68  \cr
DFT-D2$^c$ & 619.24 & 527.44 & 526.40 & 524.25 & 523.73 \cr
DFT-D3$^d$ & 620.56 & 519.68 & {\bf 518.76} & 517.73 & 518.68 \cr
\hline
\end{tabular}
\caption[]{24 atom cell volume with different combinations of van der Waals correction (VDW) and $\Gamma$-centered k space mesh, where $\Gamma$ denotes that only the gamma point is used. van der Waals dispersion corrections are: $^a$ Tkachenko-Scheffler\cite{tkatchenko2009accurate}; $^b$ TS with Hirschfeld Iteration\cite{bucko2013improved}; $^c$ Grimme D2\cite{grimme2006semiempirical}; $^d$ Grimme D3\cite{grimme2010consistent}.}
\label{kmesh_and_vdw}
\end{table}

Configurations for the 96 atom system were obtained below and above the CDW phase transition from AIMD simulations at
120 K and 300 K, respectively. The system was first equilibrated at constant temperature and pressure (the so-called NpT
ensemble) using VASP's implementation of the Parinello-Rahman
algorithm,\cite{parrinello_crystal_1980,parrinello_polymorphic_1981} which allows the cell volume and shape to
fluctuate, and an applied pressure of 1 atmosphere. The NpT runs were used to determine average lattice parameters which
were then used in subsequent runs at constant temperature and volume (NVT ensemble). Configurations were sampled at intervals
of 2 ps from the NVT runs and used to compute the thermally-averaged electronic structure of the \tise system above and 
below the CDW transition temperature.

{\it Determining lattice parameters from NpT runs}: At both 120 K and 300 K, we found that the hexagonal lattice
structure was preserved, with fluctuations in the 120 degree angle between the two lattice vectors in the \textit{xy}
plane ($<$0.3$^\circ$), and a smaller ($<$0.2$^\circ$) fluctuation between the two in-plane lattice vectors and the
out-of-plane lattice vector. At 300 K, we ran NpT dynamics for 32\,ps and observed that the lattice parameters
fluctuated but did not drift over this timescale, with average lattice vector lengths of approximately 14.19 \AA, 14.14
\AA, and 12.07 \AA, with root-mean-squared fluctuations of 0.03, 0.06 and 0.09 \AA, respectively. These lattice dimensions can be
related to the dimensions of the minimum sized CDW lattice by halving the first two ($\sim$14 \AA) values. The two \textit{xy}
lattice vectors can be taken, within the fluctuations, to be the same so we average their values to produce the average
300\,K lattice vectors given in Table~\ref{lattice_vec_table}. At 120\,K, we ran NpT dynamics for 22\,ps and observed a gradual increase in the
\textit{z}-oriented lattice vector over the first 12\,ps. We therefore averaged only over the final 10\,ps to find average
lattice vector lengths of 14.13\,\AA, 14.12\,\AA, and 12.14\,\AA, all with root-mean-squared fluctuations of
0.02\,\AA. We again averaged the lengths of the two vectors in the \textit{xy} plane to produce the average
120\,K {\it a} lattice constant given in Table~\ref{lattice_vec_table}.

\begin{table}
\caption[]{Lattice parameters (a and c) for the different $TiSe_2$ structures studied here. Units are \AA\ and root-mean-square fluctuations are given in parentheses.}
\begin{tabular}{|c|c|c|c|}
\hline
State & a (\AA) & c (\AA) & c/a \\
\hline
$P\bar{3}c1$ (CDW) & 7.08 & 12.01 & 1.70 \\
{$P\bar{3}m1$} & 7.06 & 12.01 & 1.71 \\
120 K & 7.06 (0.02) & 12.14 (0.02) & 1.71 (0.01) \\
300 K & 7.08 (0.05) & 12.07(0.05) & 1.70 (0.01)\\
\hline
\end{tabular}
\label{lattice_vec_table}
\end{table}

\section*{Summary}

Using a combination of the QS\textit{GW} and QS$G\hat{W}$ approximations with AIMD simulations, TiSe\textsubscript{2} is
shown to be a band insulator in both the low-temperature and high-temperature phases, because of symmetry lowering
deformations of the {$P\bar{3}m1$} structure.  We show explicitly in an \textit{ab initio} framework, that excitons play
a negligible role, ruling out the common assertion that TiSe\textsubscript{2} is an excitonic insulator.  The gap forms
in the low-temperature CDW phase because of level repulsions between Ti-$3d$ states at L coupling to Se-$4p$ states at
$\Gamma$ arising from matrix elements that appear by Umklapp processes from symmetry-breaking.  A gap appears in the
nominal {$P\bar{3}m1$} phase for similar reasons: but now the symmetry-breaking is dynamical, from instantaneous
deformations about the {$P\bar{3}m1$} structure.  TiSe\textsubscript{2} offers a rare instance of topological changes to
the Fermi surface originating from lattice fluctuations.

The literature on TiSe\textsubscript{2} has been confusing and contradictory because it is difficult to account for all
competing phenomena that are somewhat unique to this system.  On the scale of the small bandgap, slight perturbations
such as small nuclear deformations or approximations to the charge density, play a large role.  It is only possible to
disentangle the various effects with an exceptionally high fidelity description of the electronic structure, which the
QS\textit{GW} (QS$G\hat{W}$) approximations provide.

\section*{Acknowledgements}

This work was authored by the National Renewable Energy Laboratory, operated by Alliance for Sustainable Energy, LLC,
for the U.S. Department of Energy (DOE) under Contract No. DE-AC36-08GO28308, funding from Office of Science, Basic
Energy Sciences, Division of Materials. We acknowledge the use of the National Energy Research Scientific Computing
Center, under Contract No. DE-AC02-05CH11231 using NERSC award BES-ERCAP0021783 and we also acknowledge that a portion
of the research was performed using computational resources sponsored by the Department of Energy's Office of Energy
Efficiency and Renewable Energy and located at the National Renewable Energy Laboratory.


\bibliography{TiSe2-2023_v2.bbl}

\begin{thebibliography}{54}%
\makeatletter
\providecommand \@ifxundefined [1]{%
 \@ifx{#1\undefined}
}%
\providecommand \@ifnum [1]{%
 \ifnum #1\expandafter \@firstoftwo
 \else \expandafter \@secondoftwo
 \fi
}%
\providecommand \@ifx [1]{%
 \ifx #1\expandafter \@firstoftwo
 \else \expandafter \@secondoftwo
 \fi
}%
\providecommand \natexlab [1]{#1}%
\providecommand \enquote  [1]{``#1''}%
\providecommand \bibnamefont  [1]{#1}%
\providecommand \bibfnamefont [1]{#1}%
\providecommand \citenamefont [1]{#1}%
\providecommand \href@noop [0]{\@secondoftwo}%
\providecommand \href [0]{\begingroup \@sanitize@url \@href}%
\providecommand \@href[1]{\@@startlink{#1}\@@href}%
\providecommand \@@href[1]{\endgroup#1\@@endlink}%
\providecommand \@sanitize@url [0]{\catcode `\\12\catcode `\$12\catcode
  `\&12\catcode `\#12\catcode `\^12\catcode `\_12\catcode `\%12\relax}%
\providecommand \@@startlink[1]{}%
\providecommand \@@endlink[0]{}%
\providecommand \url  [0]{\begingroup\@sanitize@url \@url }%
\providecommand \@url [1]{\endgroup\@href {#1}{\urlprefix }}%
\providecommand \urlprefix  [0]{URL }%
\providecommand \Eprint [0]{\href }%
\providecommand \doibase [0]{http://dx.doi.org/}%
\providecommand \selectlanguage [0]{\@gobble}%
\providecommand \bibinfo  [0]{\@secondoftwo}%
\providecommand \bibfield  [0]{\@secondoftwo}%
\providecommand \translation [1]{[#1]}%
\providecommand \BibitemOpen [0]{}%
\providecommand \bibitemStop [0]{}%
\providecommand \bibitemNoStop [0]{.\EOS\space}%
\providecommand \EOS [0]{\spacefactor3000\relax}%
\providecommand \BibitemShut  [1]{\csname bibitem#1\endcsname}%
\let\auto@bib@innerbib\@empty
\bibitem [{\citenamefont {Salvo}\ \emph {et~al.}(1976)\citenamefont {Salvo},
  \citenamefont {Moncton},\ and\ \citenamefont {Waszczak}}]{DiSalvo}%
  \BibitemOpen
  \bibfield  {author} {\bibinfo {author} {\bibfnamefont {F.~J.~D.}\
  \bibnamefont {Salvo}}, \bibinfo {author} {\bibfnamefont {D.~E.}\ \bibnamefont
  {Moncton}}, \ and\ \bibinfo {author} {\bibfnamefont {J.~V.}\ \bibnamefont
  {Waszczak}},\ }\href {https://link.aps.org/doi/10.1103/PhysRevB.14.4321}
  {\bibfield  {journal} {\bibinfo  {journal} {Phys. Rev. B}\ }\textbf {\bibinfo
  {volume} {14}},\ \bibinfo {pages} {4321} (\bibinfo {year}
  {1976})}\BibitemShut {NoStop}%
\bibitem [{\citenamefont {Holt}\ \emph {et~al.}(2001)\citenamefont {Holt},
  \citenamefont {Zschack}, \citenamefont {Hong}, \citenamefont {Chou},\ and\
  \citenamefont {Chiang}}]{Holt}%
  \BibitemOpen
  \bibfield  {author} {\bibinfo {author} {\bibfnamefont {M.}~\bibnamefont
  {Holt}}, \bibinfo {author} {\bibfnamefont {P.}~\bibnamefont {Zschack}},
  \bibinfo {author} {\bibfnamefont {H.}~\bibnamefont {Hong}}, \bibinfo {author}
  {\bibfnamefont {M.-Y.}\ \bibnamefont {Chou}}, \ and\ \bibinfo {author}
  {\bibfnamefont {T.-C.}\ \bibnamefont {Chiang}},\ }\href
  {https://link.aps.org/doi/10.1103/PhysRevLett.86.3799} {\bibfield  {journal}
  {\bibinfo  {journal} {Phys. Rev. Lett.}\ }\textbf {\bibinfo {volume} {86}},\
  \bibinfo {pages} {3799} (\bibinfo {year} {2001})}\BibitemShut {NoStop}%
\bibitem [{\citenamefont {Kusmartseva}\ \emph {et~al.}(2009)\citenamefont
  {Kusmartseva}, \citenamefont {Sipos}, \citenamefont {Berger}, \citenamefont
  {Forr\'o},\ and\ \citenamefont {Tuti\v{s}}}]{Kusmartseva09}%
  \BibitemOpen
  \bibfield  {author} {\bibinfo {author} {\bibfnamefont {A.~F.}\ \bibnamefont
  {Kusmartseva}}, \bibinfo {author} {\bibfnamefont {B.}~\bibnamefont {Sipos}},
  \bibinfo {author} {\bibfnamefont {H.}~\bibnamefont {Berger}}, \bibinfo
  {author} {\bibfnamefont {L.}~\bibnamefont {Forr\'o}}, \ and\ \bibinfo
  {author} {\bibfnamefont {E.}~\bibnamefont {Tuti\v{s}}},\ }\href
  {https://link.aps.org/doi/10.1103/PhysRevLett.103.236401} {\bibfield
  {journal} {\bibinfo  {journal} {Phys. Rev. Lett.}\ }\textbf {\bibinfo
  {volume} {103}},\ \bibinfo {pages} {236401} (\bibinfo {year}
  {2009})}\BibitemShut {NoStop}%
\bibitem [{\citenamefont {Morosan}\ \emph {et~al.}(2006)\citenamefont
  {Morosan}, \citenamefont {Zandbergen},\ and\ \citenamefont
  {Dennis}}]{Morosan06}%
  \BibitemOpen
  \bibfield  {author} {\bibinfo {author} {\bibfnamefont {E.}~\bibnamefont
  {Morosan}}, \bibinfo {author} {\bibfnamefont {H.}~\bibnamefont {Zandbergen}},
  \ and\ \bibinfo {author} {\bibfnamefont {B.~e.~a.}\ \bibnamefont {Dennis}},\
  }\href {https://doi.org/10.1038/nphys360} {\bibfield  {journal} {\bibinfo
  {journal} {Nature Phys}\ }\textbf {\bibinfo {volume} {2}},\ \bibinfo {pages}
  {544} (\bibinfo {year} {2006})}\BibitemShut {NoStop}%
\bibitem [{\citenamefont {Joe}\ \emph {et~al.}(2014)\citenamefont {Joe},
  \citenamefont {Chen},\ and\ \citenamefont {et~al.}}]{Abbamonte14}%
  \BibitemOpen
  \bibfield  {author} {\bibinfo {author} {\bibfnamefont {Y.}~\bibnamefont
  {Joe}}, \bibinfo {author} {\bibfnamefont {P.}~\bibnamefont {Chen},
  \bibfnamefont {X.~Ghaemi}}, \ and\ \bibinfo {author} {\bibnamefont
  {et~al.}},\ }\href@noop {} {\bibfield  {journal} {\bibinfo  {journal} {Nat.
  Phys.}\ }\textbf {\bibinfo {volume} {10}},\ \bibinfo {pages} {421} (\bibinfo
  {year} {2014})}\BibitemShut {NoStop}%
\bibitem [{\citenamefont {Bianco}\ \emph {et~al.}(2015)\citenamefont {Bianco},
  \citenamefont {Calandra},\ and\ \citenamefont {Mauri}}]{Mauri}%
  \BibitemOpen
  \bibfield  {author} {\bibinfo {author} {\bibfnamefont {R.}~\bibnamefont
  {Bianco}}, \bibinfo {author} {\bibfnamefont {M.}~\bibnamefont {Calandra}}, \
  and\ \bibinfo {author} {\bibfnamefont {F.}~\bibnamefont {Mauri}},\
  }\href@noop {} {\bibfield  {journal} {\bibinfo  {journal} {Phys. Rev. B}\
  }\textbf {\bibinfo {volume} {92}},\ \bibinfo {pages} {094107} (\bibinfo
  {year} {2015})}\BibitemShut {NoStop}%
\bibitem [{\citenamefont {Pick}\ \emph {et~al.}(1970)\citenamefont {Pick},
  \citenamefont {Cohen},\ and\ \citenamefont {Martin}}]{Pick70}%
  \BibitemOpen
  \bibfield  {author} {\bibinfo {author} {\bibfnamefont {R.~M.}\ \bibnamefont
  {Pick}}, \bibinfo {author} {\bibfnamefont {M.~H.}\ \bibnamefont {Cohen}}, \
  and\ \bibinfo {author} {\bibfnamefont {R.~M.}\ \bibnamefont {Martin}},\
  }\href {https://link.aps.org/doi/10.1103/PhysRevB.1.910} {\bibfield
  {journal} {\bibinfo  {journal} {Phys. Rev. B}\ }\textbf {\bibinfo {volume}
  {1}},\ \bibinfo {pages} {910} (\bibinfo {year} {1970})}\BibitemShut {NoStop}%
\bibitem [{\citenamefont {Yoshida}\ and\ \citenamefont
  {Motizuki}(1980)}]{Motizuki80}%
  \BibitemOpen
  \bibfield  {author} {\bibinfo {author} {\bibfnamefont {Y.}~\bibnamefont
  {Yoshida}}\ and\ \bibinfo {author} {\bibfnamefont {K.}~\bibnamefont
  {Motizuki}},\ }\href
  {https://www.jstage.jst.go.jp/article/jpsj1946/49/3/49_3_898/_article/-char/ja/}
  {\bibfield  {journal} {\bibinfo  {journal} {J. Phys. Soc. Jpn.}\ }\textbf
  {\bibinfo {volume} {49}},\ \bibinfo {pages} {898} (\bibinfo {year}
  {1980})}\BibitemShut {NoStop}%
\bibitem [{\citenamefont {Suzuki}\ \emph {et~al.}(1985)\citenamefont {Suzuki},
  \citenamefont {Yamamoto},\ and\ \citenamefont {Motizuki}}]{Motizuki85}%
  \BibitemOpen
  \bibfield  {author} {\bibinfo {author} {\bibfnamefont {N.}~\bibnamefont
  {Suzuki}}, \bibinfo {author} {\bibfnamefont {A.}~\bibnamefont {Yamamoto}}, \
  and\ \bibinfo {author} {\bibfnamefont {K.}~\bibnamefont {Motizuki}},\ }\href
  {https://journals.jps.jp/doi/abs/10.1143/JPSJ.54.4668} {\bibfield  {journal}
  {\bibinfo  {journal} {J. Phys. Soc. Jpn.}\ }\textbf {\bibinfo {volume}
  {54}},\ \bibinfo {pages} {4668} (\bibinfo {year} {1985})}\BibitemShut
  {NoStop}%
\bibitem [{\citenamefont {Suzuki}\ \emph {et~al.}(1987)\citenamefont {Suzuki},
  \citenamefont {Yoshiyama}, \citenamefont {Motizuki},\ and\ \citenamefont
  {Takaoka}}]{Motizuki87}%
  \BibitemOpen
  \bibfield  {author} {\bibinfo {author} {\bibfnamefont {N.}~\bibnamefont
  {Suzuki}}, \bibinfo {author} {\bibfnamefont {H.}~\bibnamefont {Yoshiyama}},
  \bibinfo {author} {\bibfnamefont {K.}~\bibnamefont {Motizuki}}, \ and\
  \bibinfo {author} {\bibfnamefont {Y.}~\bibnamefont {Takaoka}},\ }\href
  {https://doi.org/10.1016/0379-6779(87)90470-X} {\bibfield  {journal}
  {\bibinfo  {journal} {Synth. Met.}\ }\textbf {\bibinfo {volume} {19}},\
  \bibinfo {pages} {887} (\bibinfo {year} {1987})}\BibitemShut {NoStop}%
\bibitem [{\citenamefont {Traum}\ \emph {et~al.}(1978)\citenamefont {Traum},
  \citenamefont {G.Margaritondo}, \citenamefont {Smith}, \citenamefont
  {J.E.Rowe},\ and\ \citenamefont {F.J.DiSalvo}}]{Traum78}%
  \BibitemOpen
  \bibfield  {author} {\bibinfo {author} {\bibfnamefont {M.}~\bibnamefont
  {Traum}}, \bibinfo {author} {\bibnamefont {G.Margaritondo}}, \bibinfo
  {author} {\bibfnamefont {N.}~\bibnamefont {Smith}}, \bibinfo {author}
  {\bibnamefont {J.E.Rowe}}, \ and\ \bibinfo {author} {\bibnamefont
  {F.J.DiSalvo}},\ }\href {https://link.aps.org/doi/10.1103/PhysRevB.17.1836}
  {\bibfield  {journal} {\bibinfo  {journal} {Phys. Rev. B}\ }\textbf {\bibinfo
  {volume} {17}},\ \bibinfo {pages} {1836} (\bibinfo {year}
  {1978})}\BibitemShut {NoStop}%
\bibitem [{\citenamefont {Wilson}(1977)}]{Wilson77}%
  \BibitemOpen
  \bibfield  {author} {\bibinfo {author} {\bibfnamefont {J.~A.}\ \bibnamefont
  {Wilson}},\ }\href {https://doi.org/10.1016/0038-1098(77)90133-8} {\bibfield
  {journal} {\bibinfo  {journal} {Solid State Commun.}\ }\textbf {\bibinfo
  {volume} {22}},\ \bibinfo {pages} {551} (\bibinfo {year} {1977})}\BibitemShut
  {NoStop}%
\bibitem [{\citenamefont {Cercellier}\ \emph {et~al.}(2007)\citenamefont
  {Cercellier}, \citenamefont {Monney}, \citenamefont {Clerc}, \citenamefont
  {Battaglia}, \citenamefont {Despont}, \citenamefont {Garnier}, \citenamefont
  {Beck}, \citenamefont {Aebi}, \citenamefont {Patthey}, \citenamefont
  {Berger},\ and\ \citenamefont {Forro}}]{Cercellier}%
  \BibitemOpen
  \bibfield  {author} {\bibinfo {author} {\bibfnamefont {H.}~\bibnamefont
  {Cercellier}}, \bibinfo {author} {\bibfnamefont {C.}~\bibnamefont {Monney}},
  \bibinfo {author} {\bibfnamefont {F.}~\bibnamefont {Clerc}}, \bibinfo
  {author} {\bibfnamefont {C.}~\bibnamefont {Battaglia}}, \bibinfo {author}
  {\bibfnamefont {L.}~\bibnamefont {Despont}}, \bibinfo {author} {\bibfnamefont
  {M.~G.}\ \bibnamefont {Garnier}}, \bibinfo {author} {\bibfnamefont
  {H.}~\bibnamefont {Beck}}, \bibinfo {author} {\bibfnamefont {P.}~\bibnamefont
  {Aebi}}, \bibinfo {author} {\bibfnamefont {L.}~\bibnamefont {Patthey}},
  \bibinfo {author} {\bibfnamefont {H.}~\bibnamefont {Berger}}, \ and\ \bibinfo
  {author} {\bibfnamefont {L.}~\bibnamefont {Forro}},\ }\href
  {https://link.aps.org/doi/10.1103/PhysRevLett.99.146403} {\bibfield
  {journal} {\bibinfo  {journal} {Phys. Rev. Lett.}\ }\textbf {\bibinfo
  {volume} {99}},\ \bibinfo {pages} {146403} (\bibinfo {year}
  {2007})}\BibitemShut {NoStop}%
\bibitem [{\citenamefont {Zunger}\ and\ \citenamefont
  {Freeman}(1978)}]{Zunger78}%
  \BibitemOpen
  \bibfield  {author} {\bibinfo {author} {\bibfnamefont {A.}~\bibnamefont
  {Zunger}}\ and\ \bibinfo {author} {\bibfnamefont {A.~J.}\ \bibnamefont
  {Freeman}},\ }\href {https://link.aps.org/doi/10.1103/PhysRevB.17.1839}
  {\bibfield  {journal} {\bibinfo  {journal} {Phys. Rev. B}\ }\textbf {\bibinfo
  {volume} {17}},\ \bibinfo {pages} {1839} (\bibinfo {year}
  {1978})}\BibitemShut {NoStop}%
\bibitem [{\citenamefont {Stoffel}\ \emph {et~al.}(1985)\citenamefont
  {Stoffel}, \citenamefont {Kevan},\ and\ \citenamefont {Smith}}]{Stoffel85}%
  \BibitemOpen
  \bibfield  {author} {\bibinfo {author} {\bibfnamefont {N.~G.}\ \bibnamefont
  {Stoffel}}, \bibinfo {author} {\bibfnamefont {S.~D.}\ \bibnamefont {Kevan}},
  \ and\ \bibinfo {author} {\bibfnamefont {N.~V.}\ \bibnamefont {Smith}},\
  }\href {https://link.aps.org/doi/10.1103/PhysRevB.31.8049} {\bibfield
  {journal} {\bibinfo  {journal} {Phys. Rev. B}\ }\textbf {\bibinfo {volume}
  {31}},\ \bibinfo {pages} {8049} (\bibinfo {year} {1985})}\BibitemShut
  {NoStop}%
\bibitem [{\citenamefont {Bachrach}\ \emph {et~al.}(1976)\citenamefont
  {Bachrach}, \citenamefont {Skibowski},\ and\ \citenamefont
  {Brown}}]{Bachrach76}%
  \BibitemOpen
  \bibfield  {author} {\bibinfo {author} {\bibfnamefont {R.~Z.}\ \bibnamefont
  {Bachrach}}, \bibinfo {author} {\bibfnamefont {M.}~\bibnamefont {Skibowski}},
  \ and\ \bibinfo {author} {\bibfnamefont {F.~C.}\ \bibnamefont {Brown}},\
  }\href {https://link.aps.org/doi/10.1103/PhysRevLett.37.40} {\bibfield
  {journal} {\bibinfo  {journal} {Phys. Rev. Lett.}\ }\textbf {\bibinfo
  {volume} {37}},\ \bibinfo {pages} {40} (\bibinfo {year} {1976})}\BibitemShut
  {NoStop}%
\bibitem [{\citenamefont {Chen}\ \emph {et~al.}(1980)\citenamefont {Chen},
  \citenamefont {Fabian}, \citenamefont {Brown}, \citenamefont {Woo},
  \citenamefont {Davies}, \citenamefont {DeLong},\ and\ \citenamefont
  {Thompson}}]{Chen80}%
  \BibitemOpen
  \bibfield  {author} {\bibinfo {author} {\bibfnamefont {C.~H.}\ \bibnamefont
  {Chen}}, \bibinfo {author} {\bibfnamefont {W.}~\bibnamefont {Fabian}},
  \bibinfo {author} {\bibfnamefont {F.~C.}\ \bibnamefont {Brown}}, \bibinfo
  {author} {\bibfnamefont {K.~C.}\ \bibnamefont {Woo}}, \bibinfo {author}
  {\bibfnamefont {B.}~\bibnamefont {Davies}}, \bibinfo {author} {\bibfnamefont
  {B.}~\bibnamefont {DeLong}}, \ and\ \bibinfo {author} {\bibfnamefont {A.~H.}\
  \bibnamefont {Thompson}},\ }\href
  {https://link.aps.org/doi/10.1103/PhysRevB.21.615} {\bibfield  {journal}
  {\bibinfo  {journal} {Phys. Rev. B}\ }\textbf {\bibinfo {volume} {21}},\
  \bibinfo {pages} {615} (\bibinfo {year} {1980})}\BibitemShut {NoStop}%
\bibitem [{\citenamefont {Cazzaniga}\ \emph {et~al.}(2012)\citenamefont
  {Cazzaniga}, \citenamefont {Cercellier}, \citenamefont {Holzmann},
  \citenamefont {Monney}, \citenamefont {Aebi}, \citenamefont {Onida},\ and\
  \citenamefont {Olevano}}]{Cazzaniga12}%
  \BibitemOpen
  \bibfield  {author} {\bibinfo {author} {\bibfnamefont {M.}~\bibnamefont
  {Cazzaniga}}, \bibinfo {author} {\bibfnamefont {H.}~\bibnamefont
  {Cercellier}}, \bibinfo {author} {\bibfnamefont {M.}~\bibnamefont
  {Holzmann}}, \bibinfo {author} {\bibfnamefont {C.}~\bibnamefont {Monney}},
  \bibinfo {author} {\bibfnamefont {P.}~\bibnamefont {Aebi}}, \bibinfo {author}
  {\bibfnamefont {G.}~\bibnamefont {Onida}}, \ and\ \bibinfo {author}
  {\bibfnamefont {V.}~\bibnamefont {Olevano}},\ }\href
  {https://journals.aps.org/prb/abstract/10.1103/PhysRevB.85.195111} {\bibfield
   {journal} {\bibinfo  {journal} {Phys. Rev. B}\ }\textbf {\bibinfo {volume}
  {85}},\ \bibinfo {pages} {195111} (\bibinfo {year} {2012})}\BibitemShut
  {NoStop}%
\bibitem [{\citenamefont {Acharya}\ \emph {et~al.}(2021)\citenamefont
  {Acharya}, \citenamefont {Pashov}, \citenamefont {Rudenko}, \citenamefont
  {Rösner}, \citenamefont {van Schilfgaarde},\ and\ \citenamefont
  {Katsnelson}}]{Acharya21a}%
  \BibitemOpen
  \bibfield  {author} {\bibinfo {author} {\bibfnamefont {S.}~\bibnamefont
  {Acharya}}, \bibinfo {author} {\bibfnamefont {D.}~\bibnamefont {Pashov}},
  \bibinfo {author} {\bibfnamefont {A.~N.}\ \bibnamefont {Rudenko}}, \bibinfo
  {author} {\bibfnamefont {M.}~\bibnamefont {Rösner}}, \bibinfo {author}
  {\bibfnamefont {M.}~\bibnamefont {van Schilfgaarde}}, \ and\ \bibinfo
  {author} {\bibfnamefont {M.~I.}\ \bibnamefont {Katsnelson}},\ }\href
  {https://doi.org/10.1038/s41524-021-00676-5} {\bibfield  {journal} {\bibinfo
  {journal} {npj Comput. Mater}\ }\textbf {\bibinfo {volume} {7}},\ \bibinfo
  {pages} {208} (\bibinfo {year} {2021})}\BibitemShut {NoStop}%
\bibitem [{\citenamefont {van Schilfgaarde}\ \emph
  {et~al.}(2006{\natexlab{a}})\citenamefont {van Schilfgaarde}, \citenamefont
  {Kotani},\ and\ \citenamefont {Faleev}}]{mark06qsgw}%
  \BibitemOpen
  \bibfield  {author} {\bibinfo {author} {\bibfnamefont {M.}~\bibnamefont {van
  Schilfgaarde}}, \bibinfo {author} {\bibfnamefont {T.}~\bibnamefont {Kotani}},
  \ and\ \bibinfo {author} {\bibfnamefont {S.}~\bibnamefont {Faleev}},\ }\href
  {http://link.aps.org/abstract/PRL/v96/e226402} {\bibfield  {journal}
  {\bibinfo  {journal} {Phys. Rev. Lett.}\ }\textbf {\bibinfo {volume} {96}},\
  \bibinfo {pages} {226402} (\bibinfo {year} {2006}{\natexlab{a}})}\BibitemShut
  {NoStop}%
\bibitem [{\citenamefont {van Schilfgaarde}\ \emph
  {et~al.}(2006{\natexlab{b}})\citenamefont {van Schilfgaarde}, \citenamefont
  {Kotani},\ and\ \citenamefont {Faleev}}]{mark06adeq}%
  \BibitemOpen
  \bibfield  {author} {\bibinfo {author} {\bibfnamefont {M.}~\bibnamefont {van
  Schilfgaarde}}, \bibinfo {author} {\bibfnamefont {T.}~\bibnamefont {Kotani}},
  \ and\ \bibinfo {author} {\bibfnamefont {S.~V.}\ \bibnamefont {Faleev}},\
  }\href {https://journals.aps.org/prb/abstract/10.1103/PhysRevB.74.245125}
  {\bibfield  {journal} {\bibinfo  {journal} {PRB}\ }\textbf {\bibinfo {volume}
  {74}},\ \bibinfo {pages} {245125} (\bibinfo {year}
  {2006}{\natexlab{b}})}\BibitemShut {NoStop}%
\bibitem [{\citenamefont {Cunningham}\ \emph {et~al.}(2023)\citenamefont
  {Cunningham}, \citenamefont {Gr{\"u}ning}, \citenamefont {Pashov},\ and\
  \citenamefont {van Schilfgaarde}}]{Cunningham2023}%
  \BibitemOpen
  \bibfield  {author} {\bibinfo {author} {\bibfnamefont {B.}~\bibnamefont
  {Cunningham}}, \bibinfo {author} {\bibfnamefont {M.}~\bibnamefont
  {Gr{\"u}ning}}, \bibinfo {author} {\bibfnamefont {D.}~\bibnamefont {Pashov}},
  \ and\ \bibinfo {author} {\bibfnamefont {M.}~\bibnamefont {van
  Schilfgaarde}},\ }\href
  {https://journals.aps.org/prb/abstract/10.1103/PhysRevB.108.165104}
  {\bibfield  {journal} {\bibinfo  {journal} {Phys. Rev. B}\ }\textbf {\bibinfo
  {volume} {108}},\ \bibinfo {pages} {165104} (\bibinfo {year}
  {2023})}\BibitemShut {NoStop}%
\bibitem [{\citenamefont {Zacharias}\ and\ \citenamefont
  {Giustino}(2016)}]{Zacharias16}%
  \BibitemOpen
  \bibfield  {author} {\bibinfo {author} {\bibfnamefont {M.}~\bibnamefont
  {Zacharias}}\ and\ \bibinfo {author} {\bibfnamefont {F.}~\bibnamefont
  {Giustino}},\ }\href
  {https://journals.aps.org/prb/abstract/10.1103/PhysRevB.94.075125} {\bibfield
   {journal} {\bibinfo  {journal} {Phys. Rev. B}\ }\textbf {\bibinfo {volume}
  {94}},\ \bibinfo {pages} {075125} (\bibinfo {year} {2016})}\BibitemShut
  {NoStop}%
\bibitem [{\citenamefont {Etienne}\ \emph {et~al.}(2016)\citenamefont
  {Etienne}, \citenamefont {Mosconi},\ and\ \citenamefont
  {Angelis}}]{Etienne16}%
  \BibitemOpen
  \bibfield  {author} {\bibinfo {author} {\bibfnamefont {T.}~\bibnamefont
  {Etienne}}, \bibinfo {author} {\bibfnamefont {E.}~\bibnamefont {Mosconi}}, \
  and\ \bibinfo {author} {\bibfnamefont {F.~D.}\ \bibnamefont {Angelis}},\
  }\href {https://doi.org/10.1021/acs.jpclett.6b00564} {\bibfield  {journal}
  {\bibinfo  {journal} {J. Phys. Chem. Lett.}\ }\textbf {\bibinfo {volume}
  {7}},\ \bibinfo {pages} {1638} (\bibinfo {year} {2016})}\BibitemShut
  {NoStop}%
\bibitem [{\citenamefont {McKechnie}\ \emph {et~al.}(2018)\citenamefont
  {McKechnie}, \citenamefont {Frost}, \citenamefont {Pashov}, \citenamefont
  {Azarhoosh}, \citenamefont {Walsh},\ and\ \citenamefont {van
  Schilfgaarde}}]{McKechnie18}%
  \BibitemOpen
  \bibfield  {author} {\bibinfo {author} {\bibfnamefont {S.}~\bibnamefont
  {McKechnie}}, \bibinfo {author} {\bibfnamefont {J.~M.}\ \bibnamefont
  {Frost}}, \bibinfo {author} {\bibfnamefont {D.}~\bibnamefont {Pashov}},
  \bibinfo {author} {\bibfnamefont {P.}~\bibnamefont {Azarhoosh}}, \bibinfo
  {author} {\bibfnamefont {A.}~\bibnamefont {Walsh}}, \ and\ \bibinfo {author}
  {\bibfnamefont {M.}~\bibnamefont {van Schilfgaarde}},\ }\href
  {https://journals.aps.org/prb/abstract/10.1103/PhysRevB.98.085108} {\bibfield
   {journal} {\bibinfo  {journal} {PRB}\ }\textbf {\bibinfo {volume} {98}},\
  \bibinfo {pages} {085108} (\bibinfo {year} {2018})}\BibitemShut {NoStop}%
\bibitem [{\citenamefont {Watson}\ \emph {et~al.}(2019)\citenamefont {Watson},
  \citenamefont {Clark}, \citenamefont {Mazzola}, \citenamefont
  {Markovi\ifmmode~\acute{c}\else \'{c}\fi{}}, \citenamefont {Sunko},
  \citenamefont {Kim}, \citenamefont {Rossnagel},\ and\ \citenamefont
  {King}}]{PhysRevLett.122.076404}%
  \BibitemOpen
  \bibfield  {author} {\bibinfo {author} {\bibfnamefont {M.~D.}\ \bibnamefont
  {Watson}}, \bibinfo {author} {\bibfnamefont {O.~J.}\ \bibnamefont {Clark}},
  \bibinfo {author} {\bibfnamefont {F.}~\bibnamefont {Mazzola}}, \bibinfo
  {author} {\bibfnamefont {I.}~\bibnamefont {Markovi\ifmmode~\acute{c}\else
  \'{c}\fi{}}}, \bibinfo {author} {\bibfnamefont {V.}~\bibnamefont {Sunko}},
  \bibinfo {author} {\bibfnamefont {T.~K.}\ \bibnamefont {Kim}}, \bibinfo
  {author} {\bibfnamefont {K.}~\bibnamefont {Rossnagel}}, \ and\ \bibinfo
  {author} {\bibfnamefont {P.~D.~C.}\ \bibnamefont {King}},\ }\href {\doibase
  10.1103/PhysRevLett.122.076404} {\bibfield  {journal} {\bibinfo  {journal}
  {Phys. Rev. Lett.}\ }\textbf {\bibinfo {volume} {122}},\ \bibinfo {pages}
  {076404} (\bibinfo {year} {2019})}\BibitemShut {NoStop}%
\bibitem [{\citenamefont {Dargam}\ \emph {et~al.}(1997)\citenamefont {Dargam},
  \citenamefont {Capaz},\ and\ \citenamefont {Koiller}}]{Dargam97}%
  \BibitemOpen
  \bibfield  {author} {\bibinfo {author} {\bibfnamefont {T.~G.}\ \bibnamefont
  {Dargam}}, \bibinfo {author} {\bibfnamefont {R.~B.}\ \bibnamefont {Capaz}}, \
  and\ \bibinfo {author} {\bibfnamefont {B.}~\bibnamefont {Koiller}},\ }\href
  {https://link.aps.org/doi/10.1103/PhysRevB.56.9625} {\bibfield  {journal}
  {\bibinfo  {journal} {Phys. Rev. B,}\ }\textbf {\bibinfo {volume} {56}},\
  \bibinfo {pages} {9625} (\bibinfo {year} {1997})}\BibitemShut {NoStop}%
\bibitem [{\citenamefont {Medeiros}\ \emph {et~al.}(2014)\citenamefont
  {Medeiros}, \citenamefont {Stafström},\ and\ \citenamefont
  {Bj{\"o}rk}}]{Medeiros14}%
  \BibitemOpen
  \bibfield  {author} {\bibinfo {author} {\bibfnamefont {P.~V.~C.}\
  \bibnamefont {Medeiros}}, \bibinfo {author} {\bibfnamefont {S.}~\bibnamefont
  {Stafström}}, \ and\ \bibinfo {author} {\bibfnamefont {J.}~\bibnamefont
  {Bj{\"o}rk}},\ }\href
  {https://journals.aps.org/prb/abstract/10.1103/PhysRevB.89.041407} {\bibfield
   {journal} {\bibinfo  {journal} {Phys. Rev. B,}\ }\textbf {\bibinfo {volume}
  {89}},\ \bibinfo {pages} {041407} (\bibinfo {year} {2014})}\BibitemShut
  {NoStop}%
\bibitem [{\citenamefont {Monney}\ \emph {et~al.}(2010)\citenamefont {Monney},
  \citenamefont {Schwier}, \citenamefont {Garnier}, \citenamefont {Mariotti},
  \citenamefont {Didiot}, \citenamefont {Beck}, \citenamefont {Aebi},
  \citenamefont {Cercellier}, \citenamefont {Marcus}, \citenamefont
  {Battaglia}, \citenamefont {Berger},\ and\ \citenamefont
  {Titov}}]{PhysRevB.81.155104}%
  \BibitemOpen
  \bibfield  {author} {\bibinfo {author} {\bibfnamefont {C.}~\bibnamefont
  {Monney}}, \bibinfo {author} {\bibfnamefont {E.~F.}\ \bibnamefont {Schwier}},
  \bibinfo {author} {\bibfnamefont {M.~G.}\ \bibnamefont {Garnier}}, \bibinfo
  {author} {\bibfnamefont {N.}~\bibnamefont {Mariotti}}, \bibinfo {author}
  {\bibfnamefont {C.}~\bibnamefont {Didiot}}, \bibinfo {author} {\bibfnamefont
  {H.}~\bibnamefont {Beck}}, \bibinfo {author} {\bibfnamefont {P.}~\bibnamefont
  {Aebi}}, \bibinfo {author} {\bibfnamefont {H.}~\bibnamefont {Cercellier}},
  \bibinfo {author} {\bibfnamefont {J.}~\bibnamefont {Marcus}}, \bibinfo
  {author} {\bibfnamefont {C.}~\bibnamefont {Battaglia}}, \bibinfo {author}
  {\bibfnamefont {H.}~\bibnamefont {Berger}}, \ and\ \bibinfo {author}
  {\bibfnamefont {A.~N.}\ \bibnamefont {Titov}},\ }\href {\doibase
  10.1103/PhysRevB.81.155104} {\bibfield  {journal} {\bibinfo  {journal} {Phys.
  Rev. B}\ }\textbf {\bibinfo {volume} {81}},\ \bibinfo {pages} {155104}
  (\bibinfo {year} {2010})}\BibitemShut {NoStop}%
\bibitem [{\citenamefont {He}\ \emph {et~al.}(2021)\citenamefont {He},
  \citenamefont {Chen}, \citenamefont {Li}, \citenamefont {Zhao}, \citenamefont
  {Song}, \citenamefont {Yoshida}, \citenamefont {Eisaki}, \citenamefont {Wu},
  \citenamefont {Chen}, \citenamefont {Lu}, \citenamefont {Meingast},
  \citenamefont {Devereaux}, \citenamefont {Birgeneau}, \citenamefont
  {Hashimoto}, \citenamefont {Lee},\ and\ \citenamefont
  {Shen}}]{PhysRevX.11.031068}%
  \BibitemOpen
  \bibfield  {author} {\bibinfo {author} {\bibfnamefont {Y.}~\bibnamefont
  {He}}, \bibinfo {author} {\bibfnamefont {S.-D.}\ \bibnamefont {Chen}},
  \bibinfo {author} {\bibfnamefont {Z.-X.}\ \bibnamefont {Li}}, \bibinfo
  {author} {\bibfnamefont {D.}~\bibnamefont {Zhao}}, \bibinfo {author}
  {\bibfnamefont {D.}~\bibnamefont {Song}}, \bibinfo {author} {\bibfnamefont
  {Y.}~\bibnamefont {Yoshida}}, \bibinfo {author} {\bibfnamefont
  {H.}~\bibnamefont {Eisaki}}, \bibinfo {author} {\bibfnamefont
  {T.}~\bibnamefont {Wu}}, \bibinfo {author} {\bibfnamefont {X.-H.}\
  \bibnamefont {Chen}}, \bibinfo {author} {\bibfnamefont {D.-H.}\ \bibnamefont
  {Lu}}, \bibinfo {author} {\bibfnamefont {C.}~\bibnamefont {Meingast}},
  \bibinfo {author} {\bibfnamefont {T.~P.}\ \bibnamefont {Devereaux}}, \bibinfo
  {author} {\bibfnamefont {R.~J.}\ \bibnamefont {Birgeneau}}, \bibinfo {author}
  {\bibfnamefont {M.}~\bibnamefont {Hashimoto}}, \bibinfo {author}
  {\bibfnamefont {D.-H.}\ \bibnamefont {Lee}}, \ and\ \bibinfo {author}
  {\bibfnamefont {Z.-X.}\ \bibnamefont {Shen}},\ }\href {\doibase
  10.1103/PhysRevX.11.031068} {\bibfield  {journal} {\bibinfo  {journal} {Phys.
  Rev. X}\ }\textbf {\bibinfo {volume} {11}},\ \bibinfo {pages} {031068}
  (\bibinfo {year} {2021})}\BibitemShut {NoStop}%
\bibitem [{\citenamefont {Seo}\ \emph {et~al.}(2019)\citenamefont {Seo},
  \citenamefont {Choi}, \citenamefont {Kimura},\ and\ \citenamefont
  {Kwon}}]{seo2019evidence}%
  \BibitemOpen
  \bibfield  {author} {\bibinfo {author} {\bibfnamefont {Y.}~\bibnamefont
  {Seo}}, \bibinfo {author} {\bibfnamefont {W.}~\bibnamefont {Choi}}, \bibinfo
  {author} {\bibfnamefont {S.-i.}\ \bibnamefont {Kimura}}, \ and\ \bibinfo
  {author} {\bibfnamefont {Y.~S.}\ \bibnamefont {Kwon}},\ }\href@noop {}
  {\bibfield  {journal} {\bibinfo  {journal} {Scientific reports}\ }\textbf
  {\bibinfo {volume} {9}},\ \bibinfo {pages} {3987} (\bibinfo {year}
  {2019})}\BibitemShut {NoStop}%
\bibitem [{\citenamefont {Bo{\v{z}}ovi{\'c}}\ and\ \citenamefont
  {Levy}(2020)}]{bovzovic2020pre}%
  \BibitemOpen
  \bibfield  {author} {\bibinfo {author} {\bibfnamefont {I.}~\bibnamefont
  {Bo{\v{z}}ovi{\'c}}}\ and\ \bibinfo {author} {\bibfnamefont {J.}~\bibnamefont
  {Levy}},\ }\href@noop {} {\bibfield  {journal} {\bibinfo  {journal} {Nature
  Physics}\ }\textbf {\bibinfo {volume} {16}},\ \bibinfo {pages} {712}
  (\bibinfo {year} {2020})}\BibitemShut {NoStop}%
\bibitem [{\citenamefont {Maki}(1968)}]{maki1968critical}%
  \BibitemOpen
  \bibfield  {author} {\bibinfo {author} {\bibfnamefont {K.}~\bibnamefont
  {Maki}},\ }\href@noop {} {\bibfield  {journal} {\bibinfo  {journal} {Progress
  of Theoretical Physics}\ }\textbf {\bibinfo {volume} {40}},\ \bibinfo {pages}
  {193} (\bibinfo {year} {1968})}\BibitemShut {NoStop}%
\bibitem [{\citenamefont {Kang}\ \emph {et~al.}(2020)\citenamefont {Kang},
  \citenamefont {Shi}, \citenamefont {Li}, \citenamefont {Wang}, \citenamefont
  {Zhang}, \citenamefont {Zhao}, \citenamefont {Li}, \citenamefont {Song},
  \citenamefont {Zheng}, \citenamefont {Nie}, \citenamefont {Wu},\ and\
  \citenamefont {Chen}}]{PhysRevLett.125.097003}%
  \BibitemOpen
  \bibfield  {author} {\bibinfo {author} {\bibfnamefont {B.~L.}\ \bibnamefont
  {Kang}}, \bibinfo {author} {\bibfnamefont {M.~Z.}\ \bibnamefont {Shi}},
  \bibinfo {author} {\bibfnamefont {S.~J.}\ \bibnamefont {Li}}, \bibinfo
  {author} {\bibfnamefont {H.~H.}\ \bibnamefont {Wang}}, \bibinfo {author}
  {\bibfnamefont {Q.}~\bibnamefont {Zhang}}, \bibinfo {author} {\bibfnamefont
  {D.}~\bibnamefont {Zhao}}, \bibinfo {author} {\bibfnamefont {J.}~\bibnamefont
  {Li}}, \bibinfo {author} {\bibfnamefont {D.~W.}\ \bibnamefont {Song}},
  \bibinfo {author} {\bibfnamefont {L.~X.}\ \bibnamefont {Zheng}}, \bibinfo
  {author} {\bibfnamefont {L.~P.}\ \bibnamefont {Nie}}, \bibinfo {author}
  {\bibfnamefont {T.}~\bibnamefont {Wu}}, \ and\ \bibinfo {author}
  {\bibfnamefont {X.~H.}\ \bibnamefont {Chen}},\ }\href {\doibase
  10.1103/PhysRevLett.125.097003} {\bibfield  {journal} {\bibinfo  {journal}
  {Phys. Rev. Lett.}\ }\textbf {\bibinfo {volume} {125}},\ \bibinfo {pages}
  {097003} (\bibinfo {year} {2020})}\BibitemShut {NoStop}%
\bibitem [{\citenamefont {Furukawa}\ \emph {et~al.}(2023)\citenamefont
  {Furukawa}, \citenamefont {Miyagawa}, \citenamefont {Matsumoto},
  \citenamefont {Sasaki},\ and\ \citenamefont
  {Kanoda}}]{PhysRevResearch.5.023165}%
  \BibitemOpen
  \bibfield  {author} {\bibinfo {author} {\bibfnamefont {T.}~\bibnamefont
  {Furukawa}}, \bibinfo {author} {\bibfnamefont {K.}~\bibnamefont {Miyagawa}},
  \bibinfo {author} {\bibfnamefont {M.}~\bibnamefont {Matsumoto}}, \bibinfo
  {author} {\bibfnamefont {T.}~\bibnamefont {Sasaki}}, \ and\ \bibinfo {author}
  {\bibfnamefont {K.}~\bibnamefont {Kanoda}},\ }\href {\doibase
  10.1103/PhysRevResearch.5.023165} {\bibfield  {journal} {\bibinfo  {journal}
  {Phys. Rev. Res.}\ }\textbf {\bibinfo {volume} {5}},\ \bibinfo {pages}
  {023165} (\bibinfo {year} {2023})}\BibitemShut {NoStop}%
\bibitem [{\citenamefont {Feng}\ \emph {et~al.}(2012)\citenamefont {Feng},
  \citenamefont {Wang}, \citenamefont {Jaramillo}, \citenamefont {Van~Wezel},
  \citenamefont {Haravifard}, \citenamefont {Srajer}, \citenamefont {Liu},
  \citenamefont {Xu}, \citenamefont {Littlewood},\ and\ \citenamefont
  {Rosenbaum}}]{feng2012order}%
  \BibitemOpen
  \bibfield  {author} {\bibinfo {author} {\bibfnamefont {Y.}~\bibnamefont
  {Feng}}, \bibinfo {author} {\bibfnamefont {J.}~\bibnamefont {Wang}}, \bibinfo
  {author} {\bibfnamefont {R.}~\bibnamefont {Jaramillo}}, \bibinfo {author}
  {\bibfnamefont {J.}~\bibnamefont {Van~Wezel}}, \bibinfo {author}
  {\bibfnamefont {S.}~\bibnamefont {Haravifard}}, \bibinfo {author}
  {\bibfnamefont {G.}~\bibnamefont {Srajer}}, \bibinfo {author} {\bibfnamefont
  {Y.}~\bibnamefont {Liu}}, \bibinfo {author} {\bibfnamefont {Z.-A.}\
  \bibnamefont {Xu}}, \bibinfo {author} {\bibfnamefont {P.}~\bibnamefont
  {Littlewood}}, \ and\ \bibinfo {author} {\bibfnamefont {T.}~\bibnamefont
  {Rosenbaum}},\ }\href@noop {} {\bibfield  {journal} {\bibinfo  {journal}
  {Proceedings of the National Academy of Sciences}\ }\textbf {\bibinfo
  {volume} {109}},\ \bibinfo {pages} {7224} (\bibinfo {year}
  {2012})}\BibitemShut {NoStop}%
\bibitem [{\citenamefont {Arpaia}\ \emph {et~al.}(2019)\citenamefont {Arpaia},
  \citenamefont {Caprara}, \citenamefont {Fumagalli}, \citenamefont
  {De~Vecchi}, \citenamefont {Peng}, \citenamefont {Andersson}, \citenamefont
  {Betto}, \citenamefont {De~Luca}, \citenamefont {Brookes}, \citenamefont
  {Lombardi} \emph {et~al.}}]{arpaia2019dynamical}%
  \BibitemOpen
  \bibfield  {author} {\bibinfo {author} {\bibfnamefont {R.}~\bibnamefont
  {Arpaia}}, \bibinfo {author} {\bibfnamefont {S.}~\bibnamefont {Caprara}},
  \bibinfo {author} {\bibfnamefont {R.}~\bibnamefont {Fumagalli}}, \bibinfo
  {author} {\bibfnamefont {G.}~\bibnamefont {De~Vecchi}}, \bibinfo {author}
  {\bibfnamefont {Y.}~\bibnamefont {Peng}}, \bibinfo {author} {\bibfnamefont
  {E.}~\bibnamefont {Andersson}}, \bibinfo {author} {\bibfnamefont
  {D.}~\bibnamefont {Betto}}, \bibinfo {author} {\bibfnamefont
  {G.}~\bibnamefont {De~Luca}}, \bibinfo {author} {\bibfnamefont
  {N.}~\bibnamefont {Brookes}}, \bibinfo {author} {\bibfnamefont
  {F.}~\bibnamefont {Lombardi}},  \emph {et~al.},\ }\href@noop {} {\bibfield
  {journal} {\bibinfo  {journal} {Science}\ }\textbf {\bibinfo {volume}
  {365}},\ \bibinfo {pages} {906} (\bibinfo {year} {2019})}\BibitemShut
  {NoStop}%
\bibitem [{\citenamefont {V.~Borzenets}\ \emph {et~al.}(2020)\citenamefont
  {V.~Borzenets}, \citenamefont {Shim}, \citenamefont {Chen}, \citenamefont
  {Ludwig}, \citenamefont {Wieck}, \citenamefont {Tarucha}, \citenamefont
  {Sim},\ and\ \citenamefont {Yamamoto}}]{v2020observation}%
  \BibitemOpen
  \bibfield  {author} {\bibinfo {author} {\bibfnamefont {I.}~\bibnamefont
  {V.~Borzenets}}, \bibinfo {author} {\bibfnamefont {J.}~\bibnamefont {Shim}},
  \bibinfo {author} {\bibfnamefont {J.~C.}\ \bibnamefont {Chen}}, \bibinfo
  {author} {\bibfnamefont {A.}~\bibnamefont {Ludwig}}, \bibinfo {author}
  {\bibfnamefont {A.~D.}\ \bibnamefont {Wieck}}, \bibinfo {author}
  {\bibfnamefont {S.}~\bibnamefont {Tarucha}}, \bibinfo {author} {\bibfnamefont
  {H.-S.}\ \bibnamefont {Sim}}, \ and\ \bibinfo {author} {\bibfnamefont
  {M.}~\bibnamefont {Yamamoto}},\ }\href@noop {} {\bibfield  {journal}
  {\bibinfo  {journal} {Nature}\ }\textbf {\bibinfo {volume} {579}},\ \bibinfo
  {pages} {210} (\bibinfo {year} {2020})}\BibitemShut {NoStop}%
\bibitem [{\citenamefont {Kotani}\ \emph {et~al.}(2007)\citenamefont {Kotani},
  \citenamefont {van Schilfgaarde},\ and\ \citenamefont {Faleev}}]{Kotani07}%
  \BibitemOpen
  \bibfield  {author} {\bibinfo {author} {\bibfnamefont {T.}~\bibnamefont
  {Kotani}}, \bibinfo {author} {\bibfnamefont {M.}~\bibnamefont {van
  Schilfgaarde}}, \ and\ \bibinfo {author} {\bibfnamefont {S.~V.}\ \bibnamefont
  {Faleev}},\ }\href
  {https://journals.aps.org/prb/abstract/10.1103/PhysRevB.76.165106} {\bibfield
   {journal} {\bibinfo  {journal} {PRB}\ }\textbf {\bibinfo {volume} {76}},\
  \bibinfo {pages} {165106} (\bibinfo {year} {2007})}\BibitemShut {NoStop}%
\bibitem [{que()}]{questaal_web}%
  \BibitemOpen
  \href@noop {} {}\bibinfo {howpublished} {\url{https://www.questaal.org}},\
  \bibinfo {note} {{Questaal code website}}\BibitemShut {NoStop}%
\bibitem [{\citenamefont {Pashov}\ \emph {et~al.}(2020)\citenamefont {Pashov},
  \citenamefont {Acharya}, \citenamefont {Lambrecht}, \citenamefont {Jackson},
  \citenamefont {Belashchenko}, \citenamefont {Chantis}, \citenamefont
  {Jamet},\ and\ \citenamefont {van Schilfgaarde}}]{questaal_paper}%
  \BibitemOpen
  \bibfield  {author} {\bibinfo {author} {\bibfnamefont {D.}~\bibnamefont
  {Pashov}}, \bibinfo {author} {\bibfnamefont {S.}~\bibnamefont {Acharya}},
  \bibinfo {author} {\bibfnamefont {W.~R.~L.}\ \bibnamefont {Lambrecht}},
  \bibinfo {author} {\bibfnamefont {J.}~\bibnamefont {Jackson}}, \bibinfo
  {author} {\bibfnamefont {K.~D.}\ \bibnamefont {Belashchenko}}, \bibinfo
  {author} {\bibfnamefont {A.}~\bibnamefont {Chantis}}, \bibinfo {author}
  {\bibfnamefont {F.}~\bibnamefont {Jamet}}, \ and\ \bibinfo {author}
  {\bibfnamefont {M.}~\bibnamefont {van Schilfgaarde}},\ }\href
  {https://www.sciencedirect.com/science/article/pii/S0010465519303868?via%3Dihub}
  {\bibfield  {journal} {\bibinfo  {journal} {Comp. Phys. Comm.}\ }\textbf
  {\bibinfo {volume} {249}},\ \bibinfo {pages} {107065} (\bibinfo {year}
  {2020})}\BibitemShut {NoStop}%
\bibitem [{\citenamefont {Kresse}\ and\ \citenamefont
  {Hafner}(1993)}]{kresse1993ab}%
  \BibitemOpen
  \bibfield  {author} {\bibinfo {author} {\bibfnamefont {G.}~\bibnamefont
  {Kresse}}\ and\ \bibinfo {author} {\bibfnamefont {J.}~\bibnamefont
  {Hafner}},\ }\href@noop {} {\bibfield  {journal} {\bibinfo  {journal}
  {Physical Review B}\ }\textbf {\bibinfo {volume} {47}},\ \bibinfo {pages}
  {558} (\bibinfo {year} {1993})}\BibitemShut {NoStop}%
\bibitem [{\citenamefont {Kresse}\ and\ \citenamefont
  {Hafner}(1994)}]{kresse1994ab}%
  \BibitemOpen
  \bibfield  {author} {\bibinfo {author} {\bibfnamefont {G.}~\bibnamefont
  {Kresse}}\ and\ \bibinfo {author} {\bibfnamefont {J.}~\bibnamefont
  {Hafner}},\ }\href@noop {} {\bibfield  {journal} {\bibinfo  {journal}
  {Physical Review B}\ }\textbf {\bibinfo {volume} {49}},\ \bibinfo {pages}
  {14251} (\bibinfo {year} {1994})}\BibitemShut {NoStop}%
\bibitem [{\citenamefont {Kresse}\ and\ \citenamefont
  {Furthm{\"u}ller}(1996{\natexlab{a}})}]{kresse1996efficiency}%
  \BibitemOpen
  \bibfield  {author} {\bibinfo {author} {\bibfnamefont {G.}~\bibnamefont
  {Kresse}}\ and\ \bibinfo {author} {\bibfnamefont {J.}~\bibnamefont
  {Furthm{\"u}ller}},\ }\href@noop {} {\bibfield  {journal} {\bibinfo
  {journal} {Computational Materials Science}\ }\textbf {\bibinfo {volume}
  {6}},\ \bibinfo {pages} {15} (\bibinfo {year}
  {1996}{\natexlab{a}})}\BibitemShut {NoStop}%
\bibitem [{\citenamefont {Kresse}\ and\ \citenamefont
  {Furthm{\"u}ller}(1996{\natexlab{b}})}]{kresse1996efficient}%
  \BibitemOpen
  \bibfield  {author} {\bibinfo {author} {\bibfnamefont {G.}~\bibnamefont
  {Kresse}}\ and\ \bibinfo {author} {\bibfnamefont {J.}~\bibnamefont
  {Furthm{\"u}ller}},\ }\href@noop {} {\bibfield  {journal} {\bibinfo
  {journal} {Physical Review B}\ }\textbf {\bibinfo {volume} {54}},\ \bibinfo
  {pages} {11169} (\bibinfo {year} {1996}{\natexlab{b}})}\BibitemShut {NoStop}%
\bibitem [{\citenamefont {Perdew}\ \emph {et~al.}(1996)\citenamefont {Perdew},
  \citenamefont {Burke},\ and\ \citenamefont
  {Ernzerhof}}]{perdew1996generalized}%
  \BibitemOpen
  \bibfield  {author} {\bibinfo {author} {\bibfnamefont {J.~P.}\ \bibnamefont
  {Perdew}}, \bibinfo {author} {\bibfnamefont {K.}~\bibnamefont {Burke}}, \
  and\ \bibinfo {author} {\bibfnamefont {M.}~\bibnamefont {Ernzerhof}},\
  }\href@noop {} {\bibfield  {journal} {\bibinfo  {journal} {Physical Review
  Letters}\ }\textbf {\bibinfo {volume} {77}},\ \bibinfo {pages} {3865}
  (\bibinfo {year} {1996})}\BibitemShut {NoStop}%
\bibitem [{\citenamefont {Bl{\"o}chl}(1994)}]{blochl1994projector}%
  \BibitemOpen
  \bibfield  {author} {\bibinfo {author} {\bibfnamefont {P.~E.}\ \bibnamefont
  {Bl{\"o}chl}},\ }\href@noop {} {\bibfield  {journal} {\bibinfo  {journal}
  {Physical Review B}\ }\textbf {\bibinfo {volume} {50}},\ \bibinfo {pages}
  {17953} (\bibinfo {year} {1994})}\BibitemShut {NoStop}%
\bibitem [{\citenamefont {Kresse}\ and\ \citenamefont
  {Joubert}(1999)}]{kresse1999ultrasoft}%
  \BibitemOpen
  \bibfield  {author} {\bibinfo {author} {\bibfnamefont {G.}~\bibnamefont
  {Kresse}}\ and\ \bibinfo {author} {\bibfnamefont {D.}~\bibnamefont
  {Joubert}},\ }\href@noop {} {\bibfield  {journal} {\bibinfo  {journal}
  {Physical Review b}\ }\textbf {\bibinfo {volume} {59}},\ \bibinfo {pages}
  {1758} (\bibinfo {year} {1999})}\BibitemShut {NoStop}%
\bibitem [{\citenamefont {Grimme}\ \emph {et~al.}(2010)\citenamefont {Grimme},
  \citenamefont {Antony}, \citenamefont {Ehrlich},\ and\ \citenamefont
  {Krieg}}]{grimme2010consistent}%
  \BibitemOpen
  \bibfield  {author} {\bibinfo {author} {\bibfnamefont {S.}~\bibnamefont
  {Grimme}}, \bibinfo {author} {\bibfnamefont {J.}~\bibnamefont {Antony}},
  \bibinfo {author} {\bibfnamefont {S.}~\bibnamefont {Ehrlich}}, \ and\
  \bibinfo {author} {\bibfnamefont {H.}~\bibnamefont {Krieg}},\ }\href@noop {}
  {\bibfield  {journal} {\bibinfo  {journal} {Journal of Chemical Physics}\
  }\textbf {\bibinfo {volume} {132}},\ \bibinfo {pages} {154104} (\bibinfo
  {year} {2010})}\BibitemShut {NoStop}%
\bibitem [{\citenamefont {Tkatchenko}\ and\ \citenamefont
  {Scheffler}(2009)}]{tkatchenko2009accurate}%
  \BibitemOpen
  \bibfield  {author} {\bibinfo {author} {\bibfnamefont {A.}~\bibnamefont
  {Tkatchenko}}\ and\ \bibinfo {author} {\bibfnamefont {M.}~\bibnamefont
  {Scheffler}},\ }\href@noop {} {\bibfield  {journal} {\bibinfo  {journal}
  {Physical Review Letters}\ }\textbf {\bibinfo {volume} {102}},\ \bibinfo
  {pages} {073005} (\bibinfo {year} {2009})}\BibitemShut {NoStop}%
\bibitem [{\citenamefont {Bucko}\ \emph {et~al.}(2013)\citenamefont {Bucko},
  \citenamefont {Lebegue}, \citenamefont {Hafner},\ and\ \citenamefont
  {Angyan}}]{bucko2013improved}%
  \BibitemOpen
  \bibfield  {author} {\bibinfo {author} {\bibfnamefont {T.}~\bibnamefont
  {Bucko}}, \bibinfo {author} {\bibfnamefont {S.}~\bibnamefont {Lebegue}},
  \bibinfo {author} {\bibfnamefont {J.}~\bibnamefont {Hafner}}, \ and\ \bibinfo
  {author} {\bibfnamefont {J.~G.}\ \bibnamefont {Angyan}},\ }\href@noop {}
  {\bibfield  {journal} {\bibinfo  {journal} {Journal of Chemical Theory and
  Computation}\ }\textbf {\bibinfo {volume} {9}},\ \bibinfo {pages} {4293}
  (\bibinfo {year} {2013})}\BibitemShut {NoStop}%
\bibitem [{\citenamefont {Grimme}(2006)}]{grimme2006semiempirical}%
  \BibitemOpen
  \bibfield  {author} {\bibinfo {author} {\bibfnamefont {S.}~\bibnamefont
  {Grimme}},\ }\href@noop {} {\bibfield  {journal} {\bibinfo  {journal}
  {Journal of Computational Chemistry}\ }\textbf {\bibinfo {volume} {27}},\
  \bibinfo {pages} {1787} (\bibinfo {year} {2006})}\BibitemShut {NoStop}%
\bibitem [{\citenamefont {Parrinello}\ and\ \citenamefont
  {Rahman}(1980)}]{parrinello_crystal_1980}%
  \BibitemOpen
  \bibfield  {author} {\bibinfo {author} {\bibfnamefont {M.}~\bibnamefont
  {Parrinello}}\ and\ \bibinfo {author} {\bibfnamefont {A.}~\bibnamefont
  {Rahman}},\ }\href {\doibase 10.1103/PhysRevLett.45.1196} {\bibfield
  {journal} {\bibinfo  {journal} {Physical Review Letters}\ }\textbf {\bibinfo
  {volume} {45}},\ \bibinfo {pages} {1196} (\bibinfo {year} {1980})},\ \bibinfo
  {note} {publisher: American Physical Society}\BibitemShut {NoStop}%
\bibitem [{\citenamefont {Parrinello}\ and\ \citenamefont
  {Rahman}(1981)}]{parrinello_polymorphic_1981}%
  \BibitemOpen
  \bibfield  {author} {\bibinfo {author} {\bibfnamefont {M.}~\bibnamefont
  {Parrinello}}\ and\ \bibinfo {author} {\bibfnamefont {A.}~\bibnamefont
  {Rahman}},\ }\href {\doibase 10.1063/1.328693} {\bibfield  {journal}
  {\bibinfo  {journal} {Journal of Applied Physics}\ }\textbf {\bibinfo
  {volume} {52}},\ \bibinfo {pages} {7182} (\bibinfo {year}
  {1981})}\BibitemShut {NoStop}%
\end{thebibliography}%

\end{document}